\documentclass[10pt]{article}
\usepackage{amsmath, amssymb,bbm}

\begin{document}
%%%%%%%%%%%%%%%%%%%%%%%%%%%%%%%%%%%%%%%%%%%%%%%%%%%%%%%%%%%%%
%\begin{titlepage}
 \rightline{LTH-989} 
%\rightline{hep-th/yymmnnn}  
%\rightline{draft: \today}  
\vskip 1.5 true cm  
\begin{center}  
{\large Non-extremal and non-BPS extremal five-dimensional black 
strings from generalized special real geometry}\\[.5em]
\vskip 1.0 true cm   
{P. Dempster$^1$ and T. Mohaupt$^{1}$} \\[3pt] 
$^1${Department of Mathematical Sciences\\ 
University of Liverpool\\
Peach Street \\
Liverpool L69 7ZL, UK\\  
P.Dempster@liv.ac.uk, Thomas.Mohaupt@liv.ac.uk \\[1em]  
%$^2${   
%Second Address  \\
%e-mail}
}
%[1em] 
%\today
October 18, 2013
%January 26, 2011 
\end{center}  
\vskip 1.0 true cm  
%%%%%%%%%%%%%%%%%%%%%%%%%%%%%%%%%%%%%%%%%%%%%%%%%%%%%%%%  
\baselineskip=18pt  
\begin{abstract}  
\noindent  

We construct non-extremal as well as extremal 
black string solutions in minimal five-dimensional supergravity
coupled to vector multiplets using dimensional reduction
to three Euclidean dimensions. Our method does not assume that
the scalar manifold is a symmetric space, and applies as well to a class 
of non-supersymmetric theories governed by a generalization of
special real geometry. We find that five-dimensional black string solutions 
correspond to geodesics in a specific totally geodesic 
para-K\"ahler submanifold
of the scalar manifold of the dimensionally reduced theory, and 
identify the subset of geodesics that corresponds to regular
black string solutions in five dimensions. BPS and non-BPS
extremal solutions are distinguished by whether the corresponding
geodesics are along the eigendirections of the para-complex 
structure or not, a characterization which carries over to non-supersymmetric
theories. 
For non-extremal black strings the values of the scalars at the
outer and inner horizon are not independent integration 
constants but determined by certain functions of the charges 
and moduli. 
By lifting solutions from three to four dimensions we obtain
non-extremal versions of small black holes, and find that while
the outer horizon takes finite size, the inner horizon is still
degenerate.

\end{abstract}

%\end{titlepage} 

\newpage
 
\tableofcontents

%%%%%%%%%%%%%%%%%%%%%%%%%%%%%%%%%%%%%%%%%%%%%%%%%%%%%%%%%%%%

\section{Introduction}

Black holes provide an important testing ground for ideas of quantum
gravity. In the context of string theory and supergravity BPS
solutions have been studied extensively since the discovery
of the attractor mechanism \cite{Ferrara:1995ih} and
of the quantitative matching between microscopic and macroscopic
entropy 
\cite{Strominger:1996sh,Maldacena:1997de,Vafa:1997gr,LopesCardoso:1998wt}.
It was realized early that many macroscopic features of 
BPS black holes, in particular the attractor mechanism, do not
strongly depend on supersymmetry and can be understood as a 
consequence of the field equations \cite{Ferrara:1997tw}.
More recently the study of non-BPS solutions has received 
increasing attention starting with 
\cite{Goldstein:2005hq,Tripathy:2005qp}, and the attractor
mechanism for general extremal black holes has been formulated
using the entropy function formalism \cite{Sen:2005wa}. The 
knowledge of non-extremal solutions is more limited and less
systematic, although many examples of non-extremal black hole
and black brane solutions in higher dimensions and in 
compactified solutions have been known for quite some time
\cite{MyersPerry,Gibbons:1987ps,Garfinkle:1990qj,Horowitz:1991cd,
Cvetic:1995dn,Cvetic:1995kv}. More recently it has been
observed that, like BPS and non-BPS extremal solutions, some
non-extremal solutions can be obtained by reducing the equations
of motion to first order form 
\cite{Lu:2003iv,Miller:2006ay,Garousi:2007zb,Andrianopoli:2007gt,Janssen:2007rc,Cardoso:2008gm,Perz:2008kh,Meessen:2011bd,Martin:2012bi,Meessen:2012su}. In this paper we will
further develop a complementary approach to non-BPS and non-extremal
solutions which aims at directly solving the second order field
equations using dimensional reduction and the special geometry
of supergravity theories with eight supercharges and their
time-reduced (Euclidean) versions. The special geometry of Euclidean
supergravities has been developed in \cite{Cortes:2003zd,Cortes:2005uq,Cortes:2009cs,Euc4}, and applied to  extremal five-dimensional black holes
\cite{Mohaupt:2009iq}, non-extremal five-dimensional black holes
\cite{Mohaupt:2010fk,Mohaupt:2012tu} and extremal
four-dimensional black holes \cite{Mohaupt:2011aa}. Our formalism
does not assume that the scalar target space is a symmetric space,
but aims to exploit the fact that for vector multiplets all couplings are
encoded in a single homogeneous function, which is real in
five dimensions and holomorphic in four dimensions. In five dimensions
one can consider models with a degree of homogeneity different
from three, which is the degree dictated by supersymmetry, and 
thus obtain a generalization of the special real geometry
of five-dimensional vector multiplets \cite{Gunaydin:1983bi},
which was dubbed `generalized special real geometry' in 
\cite{Mohaupt:2009iq,Mohaupt:2012tu}.
The non-supersymmetric theories covered by this formalism allow one
to make manifest which features of a supergravity theory do not
depend on supersymmetry \textit{per se}, but on certain features of 
the scalar manifolds which supersymmetric theories share with 
a larger class of theories.

The specific type of solution we investigate in this paper is
magnetically charged black string solutions, both extremal and
non-extremal, for five-dimensi\-onal supergravity and non-supersymmetric
theories described by generalized special real geometry. 
In five dimensions magnetic charges with respect to vector fields
are carried by strings, so that black strings are the `magnetic 
partners' of black holes, which only carry electric charges. For
minimal five-dimensional supergravity coupled to vector multiplets,
BPS black string solutions were constructed in 
\cite{Chamseddine:1999qs}. Like their electric BPS partners they 
exhibit attractor behaviour, and the Killing spinor equations 
give rise to generalized stabilization equations which allow one
to express solutions in terms of harmonic functions. Static multi-centred
BPS solutions can be obtained by choosing multi-centred harmonic
functions. More recently non-BPS extremal and non-extremal 
black string solutions have been found  using the FGK formalism
\cite{Meessen:2012su,Martin:2012bi}, which, following the
observations of \cite{Ferrara:1997tw}, employs an effective potential.
In this paper 
we approach the same problem using the formalism described above.
Our main interest is to understand the systematics and general
properties of solutions. One aspect is the relation between geodesic curves and totally geodesic 
submanifolds of the scalar manifold ${\cal M}_{(3)}$ of the
three-dimensional Euclidean theory, and solutions of the
original five-dimensional theory. Dimensional reduction 
reduces the problem of finding the equations of motion 
to the problem of finding harmonic maps from the reduced
three-dimensional space-`time' (in our case a Riemannian
space with positive signature) into ${\cal M}_{(3)}$ 
\cite{Breitenlohner:1987dg,Cortes:2009cs}. Solutions
can sometimes be constructed by identifying suitable totally geodesic
submanifolds $S\subset {\cal M}_{(3)}$, and then finding harmonic
maps from the reduced space-time into them. 
In terms of scalar fields corresponding to 
local coordinates, finding totally geodesic submanifolds
is equivalent to consistently truncating the equations of motion
by setting part of the scalar fields to constant values. 
Since we are interested in black strings in this paper, we
truncate out some of the degrees of freedom of the five-dimensional
theory from the start. We then show that the manifold obtained by dimensional
reduction to three dimensions
takes the form 
\[
S = N \times \mathbbm{R} \subset {\cal M}_{(3)},
\]
where $N$ is a para-K\"ahler manifold that can be identified
with the cotangent bundle $T^*M$ of a Hessian manifold $M$, which 
encodes the couplings of the original five-dimensional theory.
While we restrict ourselves
to the submanifold $S$ relevant for black strings in this paper, the
reduction of five-dimensional supergravity without and with
vector multiplets to three Euclidean dimensions will be studied
in depth in two companion papers \cite{G2,Euc4}.

%%%

Single-centred black string solutions correspond to geodesic
curves on $S$, which are space-like for non-extremal and null
for extremal solutions. In the extremal case one can also find
multi-centred solutions which correspond to totally geodesic,
totally isotropic submanifolds. However, not all geodesics
correspond to regular black string solutions, and the question of
which geodesics do is related to the question of how many independent
integration constants a general regular black string solution depends
on. We address this question using cases where solutions can be
obtained in closed form in terms of harmonic functions. While
this is always possible for BPS solutions in supergravity, and
a distinguished class of extremal solutions in non-supersymmetric
theories, dubbed `BPS-type solutions', the required decoupling
of the scalar equations does not happen automatically for non-extremal
and non-BPS extremal solutions. Similar to the case of five-dimensional
black holes discussed in \cite{Mohaupt:2010fk}, we show that
explicit non-extremal (and, as well, non-BPS extremal) solutions
can be obtained whenever the scalar metric of the reduced three-dimensional
theory admits a block decomposition and thus is compatible with a 
constant charge rotation matrix. The `best case', with a maximal
number of independent non-constant scalar fields expressible in 
terms of harmonic functions, are diagonal models, which include
the $ST^2$ and $STU$ models of supergravity and a class of 
$STU$-like models of non-supersymmetric theories. For these we find
explicit solutions, which for the $ST^2$ models have been derived
previously using the FGK formalism \cite{Martin:2012bi}.
We use these explicit solutions to investigate which geodesics
lift to regular black string solutions. It turns out that 
the necessary boundary conditions ensuring regularity at infinity
and at the horizon always reduce the number of integration constants
by a factor of $\frac{1}{2}$.  This resembles
the attractor mechanism for extremal solutions, which is indeed
recovered in the extremal limit. As in \cite{Mohaupt:2010fk}, where
the same behaviour was observed for five-dimensional black
holes, we refer to this phenomenon as the `deformed attractor
mechanism'. We add for clarification that for non-extremal solutions
the behaviour of solutions at the horizon remains dependent
on the values of the scalars at infinity, so that there is
no fixed-point behaviour in the strict sense. However, there are
no independent integration constants related to the horizon
values of the scalars. Moreover, the values of the scalars at the
outer and inner horizon depend on simple functions of the charges
and moduli which we dub `horizon charges'.

We also investigate extremal solutions, where we observe that
there exists a distinguished class of solutions which corresponds to 
null geodesic curves evolving along 
the eigendistributions (`eigendirections')
of the para-complex structure 
of $N$. This type of extremal solution exists in both supersymmetric
and non-supersymmetric theories, and in supersymmetric theories
these are precisely the BPS solutions. Therefore we refer to 
them as `BPS-type solutions'. They require certain restrictions
on the signs of the magnetic charges. In particular, for 
models where the scalar manifold is given by inequalities of the
form $h^I>0$, all magnetic charges must either be positive or
negative. A second, `non-BPS-type' of solution can be constructed
explicitly if a charge rotation matrix with certain properties
exists in the given model. Geometrically such solutions correspond
to null geodesic curves which do not evolve along the 
eigendistributions of the para-complex structure.
In supersymmetric models these solutions are extremal, but not
BPS. In  models with scalar manifolds of the form $h^I>0$, 
such solutions carry magnetic charges which are not all positive
or all negative. For a class of models which includes the $ST^2$
and $STU$ models of supergravity, as well as $STU$-like solutions
of non-supersymmetric theories, we show explicitly that charge rotation matrices
giving rise to all possible choices of signs exist. For generic
models our observation explains geometrically 
why non-BPS extremal solutions are 
harder to find than BPS solutions.

The structure of this paper is as follows. In Section 2 we briefly review
black string solutions in five-dimensional Einstein-Maxwell theory. In 
Section 3 we review the special real geometry of five-dimensional 
vector multiplets and carry out the reduction of the relevant part
of the theory to three dimensions. We observe that the target 
manifold is the product of a para-K\"ahler manifold with a 
one-dimensional factor. A short self-contained
proof of the para-K\"ahler property is relegated to Appendix \ref{AppHessianPK}. 
In Section 4 we solve the three-dimensional Einstein equations
and observe that the three-dimensional line element is universal
and coincides with the reduced line element of a five-dimensional
`Reissner-Nordstr\"om string'. In Section 5 we solve the scalar field
equations, while in Section 6 we discuss the resulting five-dimensional
non-extremal black string solutions. In Section 7 we obtain extremal
black string solutions and compare BPS and non-BPS-type solutions.
For the $ST^2$ and $STU$ models we compare our method to the FGK
formalism used in \cite{Martin:2012bi,Meessen:2012su}. In Section 9
we present the generalization to non-supersymmetric theories and 
uncover the relation between the BPS condition and eigendirections
of the para-complex structure of the scalar manifold. Our conclusions
are given in Section 10. Appendix \ref{AppHessianPK} contains a short proof that the submanifold $N \subset
N \times \mathbbm{R} = {\cal M}_{(3)}$ of the scalar manifold of
the reduced theory is para-K\"ahler.

\section{\sloppy Black strings in five-dimensional Einstein-Maxwell theory}

For reference we briefly review the basic black string solution
of five-dimensional Einstein-Maxwell theory, which might be viewed
as a variant of the four-dimensi\-onal Reissner-Nordstr\"{o}m solution.
This solution is a special example of a Reissner-Nordstr\"{o}m type 
black brane solution, which exist in various dimensions and which 
are reviewed, for example, in \cite{Martin:2012bi}. A RN (Reissner-Nordstr\"{o}m)
type black string solution has an isometry group which contains
a static time-like Killing vector field and space-like translational
Killing vector field which commute with one another and with the
transverse rotation group $SO(3)$:
\[
\mbox{Isom} \supset \mathbbm{R}_t \times \mathbbm{R}_y \times SO(3).
\]
When using adapted coordinates $(t,y,\rho,\theta,\phi)$, the line
element can be brought to the form \cite{Martin:2012bi}
\begin{equation}
\label{5dRNBS}
ds^2_{(5)} = H^{-1}(\rho) \left[ - W(\rho) dt^2 + dy^2 \right]
+ \frac{H^2(\rho)}{W(\rho)} \left[ d\rho^2 + W(\rho)\rho^2 d \Omega_{(2)}^2 \right],
\end{equation}
where $d\Omega_{(2)}^2$ is the line element of the round unit 2-sphere,
and where
\[
H = 1 + \frac{{p}}{\rho} \;,\;\;\;
W = 1 - \frac{2c}{\rho} \;.
\]
The two parameters $p,c$ are non-negative: ${p}\geq 0$,
$c\geq 0$. The solution has an outer horizon at $\rho=2c$ and
an inner horizon at $\rho=0$. To explore the region inside the
inner horizon one can choose different coordinates, see for example
\cite{Martin:2012bi}, but the coordinate system above will be convenient
later. For $c=0$ one obtains the extremal limit where both horizons
coincide, thus identifying $c$ as the non-extremality parameter.
The second parameter
$p$ is related to the magnetic charge of the black string. 
The non-vanishing component of the field strength is
\[
F_{\theta \phi} \simeq \pm {p} \sin \theta \;,
\]
which implies that the magnetic charge is $\tilde{p}=\pm p$. 
Observe that the magnetic charge can be positive or negative,
whereas the parameter ${p}$ must be non-negative. 
For negative ${p}$ the coefficients of the line element
will have additional zeroes and infinities, which correspond
to naked singularities, see for example \cite{Mayer:2003zk}.
We remark that the overall sign between the magnetic charge $\tilde{p}$
and the parameter $p$ is not determined by the field 
equations, so that choosing this sign is part of specifying the
solution.

We finally recall that black string solutions are subject to 
an extremality bound of the form 
\[
\mathcal{T} \geq \mbox{Const} |\tilde{p}| \;,
\]
where $\mathcal{T}$ is the ADM tension, see \cite{Mayer:2003zk} for more
details. 
For static BPS string solutions in five-dimensional
supergravity, this
extremality bound is implied by the BPS bound, which takes the form
\[
\mathcal{T} \geq \mbox{Const} |\mathcal{Z}_m|,
\]
where $\mathcal{Z}_m$ is the `magnetic central charge'
\cite{Antoniadis:1995vz,Chou:1997ba,Chamseddine:1999qs} of the
string. 
As for black holes, supersymmetric theories can also have extremal
solutions which are not BPS, i.e. solutions which satisfy the
extremality bound but not the BPS bound.

In the following our goal is to construct non-BPS solutions,
both non-extremal and extremal, in five-dimensional supergravity
with vector multiplets and, more generally, five-dimensional
Einstein-Vector-Scalar type theories where the couplings are 
determined by `generalized special real geometry' as defined in  
\cite{Mohaupt:2009iq,Mohaupt:2010fk,Mohaupt:2012tu}. 
As the solutions are in general non-BPS,
we need to solve the full field equations. This is done
by dimensional reduction to three space-like dimensions using the
existence of two commuting Killing vector fields corresponding
to staticity and translations along the string. We then use the
formalism of `generalized special geometry' and exploit the fact that 
all couplings are encoded in a single function, the Hesse potential.

%\section{Dimensional reduction and instanton solutions}

\section{Dimensional reduction}\label{dimred}

We begin with the action for minimal 
five-dimensional supergravity coupled to some number, $n_V^{(5)}$, of vector multiplets \cite{Gunaydin:1983bi}. In the conventions of 
\cite{Cortes:2009cs}, the bosonic part of the action takes the form
\begin{eqnarray}\label{5dAction}
S_5 &=&\int d^5x\left[\sqrt{\hat{g}}\left(\frac{\hat{R}}{2}-\frac{3}{4}
a_{ij}(h)\partial_{\hat{\mu}} h^i\partial^{\hat{\mu}} h^j
-\frac{1}{4}a_{ij}(h)\mathcal{F}^i_{\hat{\mu}\hat{\nu}}\mathcal{F}^{j|\hat{\mu}\hat{\nu}}\right)\right. \nonumber \\
& &\hspace{15mm}\left. +\frac{1}{6\sqrt{6}}c_{ijk}\epsilon^{\hat{\mu}\hat{\nu}\hat{\rho}\hat{\sigma}\hat{\lambda}}\mathcal{F}^i_{\hat{\mu}\hat{\nu}}\mathcal{F}^j_{\hat{\rho}\hat{\sigma}}\mathcal{A}^k_{\hat{\lambda}}\right].
\end{eqnarray}

Here $\hat{\mu},\hat{\nu},\ldots$ are five-dimensional Lorentz indices and $i=1,\ldots,n_V^{(5)}+1$ labels the five-dimensional gauge fields. The scalars $h^i$ are understood to satisfy the constraint
\begin{equation}\label{Hypersurface}
H(h)=c_{ijk}h^ih^jh^k=1,
\end{equation}
which defines an $n_V^{(5)}$-dimensional submanifold 
$\mathcal{H}\subset M$, where $M$ is a real manifold of
dimension $n_V^{(5)}+1$. The fields $h^i$ can be interpreted
as coordinates for $M$ and as homogeneous coordinates for 
the hypersurface ${\cal H}$.

The symmetric, positive definite tensor field $a_{ij}(h)$ 
appearing in the action (\ref{5dAction})
is obtained by taking the second derivatives 
\begin{equation}
\label{aij}
a_{ij} =\frac{\partial^2 \tilde{H}}{\partial h^i
\partial h^j}\; ,
\end{equation}
of the Hesse potential
\begin{equation}
\label{tildeH}
\tilde{H}=-\frac{1}{d}\log H.
\end{equation}
%c_{ijk} \simeq \frac{\partial^3 \hat{\mathcal{V}}}{\partial h^i 
%\partial h^j \partial h^k} \;.
%\]

The tensor $a_{ij}(h)$ defines a positive definite Hessian metric 
$ds^2_M = a_{ij} dh^i dh^j$ on $M$. One property of Hessian 
metrics which we use later is that the first derivatives
$\partial_k a_{ij}$, and therefore also the Christoffel symbols
of the first kind, are totally symmetric in all three indices. 
We will also use that the metric coefficients $a_{ij}$
are homogeneous functions of degree $-2$ with respect 
to the coordinates $h^i$. Recall that a homogeneous function 
$f(h^i)$ of degree $n$ satisfies the Euler relation 
$h^i \partial_i f = n f$. The metric coefficients $a_{ij} = \partial^2_{i,j} H$
of a metric with a Hesse potential 
$H$ that is homogeneous of 
degree $n$ are themselves homogeneous of degree $n-2$. If one
takes the Hesse potential $\tilde{H}$ to be proportional to the
logarithm of a homogeneous function $H$ (of any degree), as in 
(\ref{tildeH}), then $\tilde{H}$ itself is not a homogeneous
function. However, its
$k^{\textrm{th}}$ derivatives ($k>1$) are homogeneous functions
of degree $-k$ and, in particular, the metric coefficients of the
corresponding Hessian metric (\ref{aij}) are homogeneous of degree $-2$.

While the vector couplings are given by restricting the
tensor $a_{ij}$ to the hypersurface $H=1$,
the couplings of the physical (independent) scalars are 
given by the pullback of $a_{ij}$ to ${\cal H}$. To make
this explicit one can solve \eqref{Hypersurface} in terms
of $n_V^{(5)}$ independent scalars, which then provide
(inhomogeneous) coordinates for ${\cal H}$. For us
it is more convenient to work with the dependent scalars $h^i$
for reasons that will become clear later.

We remark that the formalism we use in the following only depends
on the fact that $H$ is a homogeneous function,
and not on the more specific condition that it is a polynomial and
has degree three. These additional conditions follow from imposing
that the theory is supersymmetric. By allowing a non-polynomial function with degree of homogeneity
different from three, one obtains a more general class of 
non-supersymmetric theories of vector and scalar fields (and possibly
fermions) coupled to gravity. The formalism of generalized special
geometry developed in \cite{Mohaupt:2009iq,Mohaupt:2010fk,Mohaupt:2012tu} 
allows one to solve the field equations within this larger class in 
precisely the same way as in supergravity. For concreteness we
will in the following focus on supergravity.
The generalization to general
homogeneous $H$ is however completely straightforward and 
will be discussed in Section \ref{Section_GSG}.

%%%

We are interested here in five-dimensional string-like solutions which are static and magnetically-charged under the gauge fields $\mathcal{A}^i_{\hat{\mu}}$. As such our solutions will admit one timelike and one spacelike isometry (along the direction of the string) and so we can use the techniques of dimensional reduction over one timelike and one spacelike direction to generate solutions.

%In particular, we impose the metric ansatz $M_5=S^1\times S^1\times M_3$ with
In particular we impose that the line element takes the form
\begin{equation}\label{MetricAnsatz}
ds_{(5)}^{2}=-\epsilon_{1}e^{2\sigma}\left(dx^{0}\right)^{2}
-\epsilon_2 e^{2\phi-\sigma}\left(dx^4\right)^2 +e^{-2\phi-\sigma}ds^2_{(3)},
\end{equation}
where the two as yet undetermined functions $\sigma$ and $\phi$ only depend
on the coordinates of the reduced three-dimensional space with as yet
undetermined line element $ds_{(3)}^2$. 
Our parametrization has been
chosen such that $\sigma$ and $\phi$ are the Kaluza-Klein scalars
of the dimensional reductions from the five-dimensional to the
four-dimensional Einstein frame, and from the four-dimensional to the
three-dimensional Einstein frame, respectively. 
The parameters
$\epsilon_{1,2}$ take the values $-1$ for reduction over a spacelike direction and $+1$ for reduction over a timelike direction\footnote{Note that in the case at hand we reduce over one timelike and one spacelike direction, so will always take $\epsilon_2=-\epsilon_1$. However, we leave the general case for convenience.}. Note that we can take either $x^0$ or $x^4$ to be timelike. The seemingly asymmetric treatment of $\lbrace x^0,x^4\rbrace$ stems from the fact that we first perform a reduction (taken to be either timelike or spacelike depending on the sign of $\epsilon_1$) over $x^0$ and \textit{then} a reduction over $x^4$.
Our parametrization allows us to postpone the decision as to whether we first
reduce over time and then over space, or vice versa. While it will turn
out that when restricting to those fields which are non-trivial for 
black string solutions this choice is not relevant, the distinction
becomes relevant when considering all fields. This will be discussed
in a separate publication \cite{G2}.

%We note that, for the solutions we will be interested in, the choice of whether one starts with the spacelike or timelike reduction is irrelevant. The analysis of the more generic case, where this choice becomes important, will appear in future work.

Furthermore, restricting ourselves to magnetic solutions leads us to impose the ansatz for the gauge fields $\mathcal{A}^i_{\hat{\mu}}$,
\begin{equation}
\label{GaugeFieldAnsatz}
\mathcal{A}^i_{\hat{\mu}}dx^{\hat{\mu}}=A^i_\mu dx^\mu,
\end{equation}
where $x^\mu$ with $\mu=1,2,3$ are coordinates transverse to the string. In other words, we set $\mathcal{A}^i_0=\mathcal{A}^i_4=0$.

For this class of solutions, the resulting three-dimensional Euclidean action is
\begin{equation}\label{BSAction}
S_3=\int d^3x\sqrt{g}\left[\frac{R}{2}-\hat{g}_{ij}(y)\partial_\mu y^i\partial^\mu y^j-(\partial\phi)^2 + e^{-2\phi-3\sigma}\hat{g}^{ij}(y)\partial_\mu s_i\partial^\mu s_j
\right],
\end{equation}
where $R$ is the three-dimensional Ricci scalar which does not give rise to local dynamics. %\\[2ex]
%
%{\bf TMM: I have started to implement the following changes of conventions:
%$\hat{g}_{ij} \rightarrow - \hat{g}_{ij}$, so that this quantity is
%positive definite, $s_i \rightarrow \sqrt{8} s_i$, to absorbe the
%factor 1/8.} \\[2ex]
%

The dynamical fields are the $2n_V^{(5)}+3$ scalar fields $(y^i,s_i,\phi)$, 
which have the following five-dimensional origin: the scalars $y^i$ encode the degrees of freedom of the original (constrained) scalars $h^i$ and the Kaluza-Klein scalar from the five-to-four reduction, $\sigma$, via
\[
y^i=6^{\frac{1}{3}}e^\sigma h^i,
\]
and are therefore unconstrained; the scalar $\phi$ arises as the Kaluza-Klein scalar in the reduction from four to three dimensions; the axions $s_i$ are obtained by dualizing the gauge fields $A^i_\mu$ after reduction to three dimensions.
Finally, using homogeneity we can express the metric $a_{ij}$ in terms 
of the rescaled fields $y^i$. Including a constant overall factor we
obtain
%we have rescaled the metric $a_{ij}$ on the original special real manifold to 
\cite{Cortes:2009cs}
\begin{equation}\label{gMetric}
\hat{g}_{ij}(y)= - \frac{3}{2}\left(\frac{(cy)_{ij}}{cyyy}-\frac{3}{2}\frac{(cyy)_i(cyy)_j}{(cyyy)^2}\right) = - \frac{1}{4} \partial^2_{y^i,y^j} 
\log \left( c_{klm} y^k y^l y^m \right).
\end{equation}

We note that this metric is Hessian, and homogeneous of degree $-2$.
For later use, we also note the identity 
$\hat{g}_{ij}(y) y^i y^j = \frac{3}{4} \hat{g}_{ij}(y)$. 

The action \eqref{BSAction} can be simplified by a further field redefinition
\[
w^i=e^{-\phi-\frac{3}{2}\sigma}y^i,\hspace{5mm} \xi=\phi-\frac{3}{2}\sigma,
\]
which gives
\begin{equation}\label{BSAction2}
S_3=\int d^3x\sqrt{g}\left[\frac{R}{2}-\hat{g}_{ij}(w)\partial_\mu w^i\partial^\mu w^j+\hat{g}^{ij}(w)\partial_\mu s_i\partial^\mu s_j -\frac{1}{4}(\partial\xi)^2
\right],
\end{equation}
where we now take $(w^i,\xi,s_i)$ to be the dynamical fields. These fields
parametrize a $(2n_V^{(5)} +3)$-dimensional 
submanifold $S$ of the full $(4n_V^{(5)} + 8)$-dimensional manifold
${\cal M}_{(3)}$ 
which is obtained if all five-dimensional degrees of freedom are 
kept and dualized into scalars. As we will show in separate 
publications \cite{G2,Euc4}, the full manifold ${\cal M}_{(3)}$ is
a para-quaternionic K\"ahler manifold. Here we restrict ourselves
to investigating the geometry of the submanifold $S$. 
The manifold $S$ is a totally geodesic submanifold of ${\cal M}_{(3)}$, 
since it is obtained by solving
the equations of motion for 
$2n_V^{(5)} + 5$ out of $4n_V^{(5)} + 8$ scalars 
by setting them to constant values. The fields which are truncated
out are (i) three out of five degrees of freedom of the five-dimensional
metric, see (\ref{MetricAnsatz}), or, equivalently, the scalars 
corresponding 
to the Kaluza-Klein vectors of the two reduction steps, and (ii) 
$2(n_V^{(5)} + 1)$ out of $3(n_V^{(5)} +1)$ 
degrees of freedom of the five-dimensional vector fields, see
(\ref{GaugeFieldAnsatz}), or, equivalently, the corresponding
three-dimensional scalars. The line element of the submanifold
$S$ takes the form
\[
ds^2_S = \hat{g}_{ij}(w) dw^i dw^j - \hat{g}^{ij}(w) ds_i ds_j 
+ \frac{1}{4} (d\xi)^2 \;.
\]

The metric on $S$ is the product of a one-dimensional factor
parametrized by $\xi$ and a $2(n_V^{(5)} +1)$-dimensional 
manifold ${N}$, which can be identified with the 
cotangent bundle of the manifold $M$ of the five-dimensional
theory, ${N} \simeq T^* M$. Moreover, since $ds^2_M = \hat{g}_{ij}(w)
dw^i dw^j$ is a Hessian metric, it follows that 
$ds^2_N = \hat{g}_{ij}(w) dw^i dw^j - \hat{g}^{ij}(w) ds_i ds_j $ is a 
para-K\"ahler metric on $N$, as we show in Appendix \ref{AppHessianPK}.

%The relation between 
%the metrics $\hat{g}_{ij}(w) dw^i dw^j$ of $M$ and 
%$\hat{g}_{ij}(w) dw^i dw^j - \hat{g}^{ij}(w) ds_i ds_j$ is 
%analgous to the situation arising for the r-map,
%which relates the scalar manifolds of five- and four-dimensional
%vector multiplets. This type of metric is discussed in detail in 
%\cite{Cortes:2005uq} and \cite{Mohaupt:2009iq}. 

%In particular,
%it was shown that if $\hat{g}_{ij}$ is a Hessian metric
%on a real manifold $M$, then, 
%$\hat{g}_{ij}(w) ( dw^i dw^j - ds^i ds^j)$ is a para-K\"ahler
%metric on the tangent bundle $TM$, while 
%$ds^2_{N} = \hat{g}_{ij}(w) dw^i dw^j -  \hat{g}^{ij}(w) ds_i ds_j$
%is the corresponding para-K\"ahler metric on the
%cotantent bundle $T^*M$. {\bf TMM: add reference or argument
%for cotangent bundle, probably Euc-II paper. 
%How are $s^i$ and $s_i$ related?}

%is para-K\"ahler. Since the metric $\hat{g}_{ij}$ provides a
%natural isomorphism between $\bar{N}=T^*M$ and $TM$, one can
%show that the metric is para-K\"ahler by verifying 
%that the associated metric 
%$ds^2_{TM} = \hat{g}_{ij}(w) dw^i dw^j -  \hat{g}^{ij}(w) ds_i ds_j$
%on $TM$ has a local para-K\"ahler potential, which is proportional
%to the Hesse potential of $\hat{g}_{ij}(w)$, see \cite{Mohaupt:2009iq}
%for details. Alternatively, one can use that the Hessian manifold
%$M$ comes equipped with a `special' connection, see \cite{Cortes:2005uq}.
%Thus the submanifold $S$ relevant for 
%black string solutions is the product of an $2(n_V^{(5)} +1)$-dimensional
%para-K\"ahler manifold with flat one-dimensional factor. 
 
We next observe that for the subsector of fields relevant
for black string solutions the parameters $\epsilon_1$ and
$\epsilon_2$ do not appear explicitly in the 
action (\ref{BSAction2}). Thus this subsector is manifestly 
insensitive to whether we first reduce over time or over
space. As we will discuss in \cite{G2}, \cite{Euc4}, this is
different when the full set of fields is considered.

For later reference, we list 
the relations between the three-dimensional fields and 
our original five-dimensional fields. Specifically,
\begin{equation}\label{MetricAnsatzTrunc}
ds^2_{(5)}=e^{\xi+2\sigma}\left[-\epsilon_1 e^{-\xi}(dx^0)^2-\epsilon_2 e^\xi (dx^4)^2\right]+e^{-2(\xi+2\sigma)}ds^2_{(3)},
\end{equation}
for the metric, and 
\begin{equation}\label{TruncFields}
h^i=e^{\xi+2\sigma}w^i, \hspace{5mm} F^i_{\mu\nu}=-\frac{1}{\sqrt{2}}\epsilon_{\mu\nu\rho}\hat{g}^{ij}(w)\partial^\rho s_j,
\end{equation}
for the remaining fields.

%\section{Solving the equations of motion}

\section{Solving the three-dimensional Einstein equations}

We now turn our attention to the three-dimensional equations of motion coming from the action \eqref{BSAction2}. The Einstein equations (after taking a trace and back-substituting) read
\begin{equation}\label{EinsteinEqn}
\frac{1}{2}R_{\mu\nu}-\hat{g}_{ij}(w)\partial_\mu w^i\partial_\nu w^j
+ \hat{g}^{ij}(w)\partial_\mu s_i\partial_\nu s_j -\frac{1}{4}\partial_\mu\xi\partial_\nu\xi=0.
\end{equation}

We will look primarily for solutions describing a single static 
black string and which therefore possess spherical symmetry in the
three-dimensional transverse space. We remark that one could dispense with spherical symmetry when considering extremal solutions, thus allowing for the possibility of multi-centred solutions. While this is not the main focus of this work,
we will come back to this point later when we discuss extremal solutions.

Any spherically symmetric line element in 3 dimensions can be
brought to the form
\begin{equation}\label{SphSymm}
ds^2_{(3)}=e^{4A(\tau)}d\tau^2 +e^{2A(\tau)}\left(d\theta^2+\sin^2\theta d\varphi^2\right),
\end{equation}
where $\tau$ is a radial coordinate \cite{Perz:2008kh}. 
Spherical symmetry of the field configuration then imposes that the scalar fields $(w^i,s_i,\xi)$ are independent of the angular coordinates $(\theta,\varphi)$.

Plugging this ansatz into \eqref{EinsteinEqn} with $m,n\neq\tau$ we find
\begin{equation}\label{Eins1}
1-e^{-2A}\ddot{A}=0,
\end{equation}
where here $\dot{X}$ denotes differentiation with respect to $\tau$. 
Multiplying \eqref{Eins1} through by $2e^{2A}\dot{A}$ we obtain
\[
\frac{d}{d\tau}(e^{2A}-\dot{A}^2)=0,
\]
which can be integrated to find
\begin{equation}
\dot{A}^2=e^{2A}+\mu,
\end{equation}
for some integration constant $\mu$. Taking the square root and multiplying through by $-e^{-A}$ gives the differential equation
\begin{equation}
\frac{d}{d\tau}e^{-A}=\sqrt{1+\mu e^{-2A}},
\end{equation}
which can be solved to find an expression for $e^{A(\tau)}$ provided we make a choice for the sign of $\mu$.
%
%{\bf TMM: Give details how separation constant $\mu = c^2$ arises.}\\[2ex]
If we choose the integration constant to be positive, $\mu = c^2 >0$,
we obtain the general solution
\begin{equation}
\label{expAtau}
e^{A(\tau)}=\frac{c}{\sinh(c\tau)},
\end{equation}
where the real constant $c$ is chosen positive, $c>0$, for
concreteness. 
We will see later that solutions with $c=0$ are
well-defined and correspond to the extremal limit, thus
identifying $c$ as the non-extremality parameter. Since (\ref{expAtau})
is manifestly invariant under $c\rightarrow -c$, we do not need to
consider $c<0$. In solutions with negative values
$\mu<0$ of the integration constant the hyperbolic function appearing 
in (\ref{expAtau}) is replaced by a trigonometric function. In this
case the `radial' coordinate $\tau$ is periodic, and such solutions
cannot lift to asymptotically flat black string solutions. We therefore
discard solutions with $\mu<0$.

With this, the three-dimensional part of the metric \eqref{SphSymm} becomes
\begin{equation}\label{3dMetric}
ds^2_{(3)}=\frac{c^4}{\sinh^4(c\tau)}d\tau^2+\frac{c^2}{\sinh^2(c\tau)}d\Omega_2^2.
\end{equation}

Returning now to the remaining equations \eqref{EinsteinEqn}, namely those with $m=n=\tau$, we obtain
\begin{equation}\label{HamConstraint}
c^2-\hat{g}_{ij}(w)\dot{w}^i\dot{w}^j+\hat{g}^{ij}(w)\dot{s}_i\dot{s}_j -\frac{1}{4}\dot{\xi}^2=0.
\end{equation}
This relation is often called the Hamiltonian constraint. If one 
imposes spherical symmetry at the level of the action and reduces
the action to one dimension, this equation no longer follows from 
the variational principle and thus has to be imposed as an additional
condition. We instead obtained it as a field equation because we
imposed spherical symmetry on the three-dimensional field equations,
and not on the action itself. 
The Hamiltonian constraint allows the following interpretation in
terms of the scalar manifold $S$. A spherically symmetric solution
corresponds to a geodesic curve $C$ on $S$, parametrized by $\tau$,
with tangent vector $(\dot{w}^i, \dot{s}_i, \dot{\xi})$. The 
Hamiltonian constraint implies that this tangent vector has constant
scalar product $\mu=c^2$ with itself. Therefore the radial coordinate
$\tau$ is an affine curve parameter.
Moreover curves with 
$\mu = c^2 >0$ are space-like while curves with $c=0$ are light-like (null).
We will see later that geodesics with $c^2>0$ satisfying appropriate
boundary conditions lift to non-extremal black string solutions, 
while geodesics with $c^2=0$ lift to extremal black string solutions.
As we have seen above, space-like geodesics ($\mu<0$) do not lift
to black string solutions.

It is useful to introduce a new radial coordinate
\begin{equation}\label{IsoCoords}
\rho=\frac{ce^{c\tau}}{\sinh(c\tau)},
\end{equation}
which no longer corresponds to an affine coordinate on the geodesic
curve $C$ on $S$. 

In terms of $\rho$ the line element \eqref{3dMetric} takes the form
\begin{equation}\label{3dMetric2}
ds^2_{(3)}=d\rho^2+W\rho^2 d\Omega_2^2,
\end{equation}
where
\begin{equation}\label{NonExtrHarm}
W:=1-\frac{2c}{\rho}=e^{-2c\tau},
\end{equation}
is harmonic in the three-dimensional transverse space.
This is exactly the same as the three-dimensional part of the
line element of the standard five-dimensional 
RN-type black string (\ref{5dRNBS}). Thus, as for five-dimensional
black holes (see for example \cite{Mohaupt:2010fk}), the geometry of this
three-dimensional part is universal and remains the same when 
the solution is deformed by allowing a non-trivial profile for
scalar fields.

We also observe\footnote{Here we anticipate that the following
discussion is not modified by the presence of non-constant 
scalar fields. This is justified by the discussion at the end
of Section 6.} 
that the range $0 < \tau < \infty$ of the
`affine' radial coordinate $\tau$ corresponds to the range
$\infty > \rho > 2c$ of the standard radial coordinate, which
covers the region between the asymptotically flat limit
$\tau \rightarrow 0 \Leftrightarrow \rho \rightarrow \infty$
and the outer horizon at $\tau \rightarrow \infty \Leftrightarrow
\rho \rightarrow 2c$.
As in \cite{Mohaupt:2010fk} one can
therefore use the coordinate $\rho$ to continue the solution 
to the region between the outer horizon at $\rho=2c$ and the
inner horizon $\rho=0$.
 Given that we used dimensional reduction 
over time it is clear that we should only expect to obtain a
solution valid up to the outer horizon, because the Killing vector
field $\partial_t$ is not time-like but space-like 
for $2c > \rho >0$. Thus in this region one would have to use
a dimensional reduction with respect to two space-like directions,
leading to a different auxiliary three-dimensional theory.

\section{\sloppy Solving the three-dimensional scalar equations of motion}\label{Instantons}

We now turn to the equations of motion for the scalar fields
$(w^i,s_i,\xi)$, which by assumption of spherical symmetry
only depend on the radial coordinate $\tau$. The equations of motion 
for these $2n_v^{(5)} +3$ fields are of second order. Therefore 
the general solution,
which is guaranteed to exist at least locally, will depend on 
$2(2n_V^{(5)} +3)$ integration constants. Geometrically, solutions
correspond to geodesic curves on $S$ and the integration constants
correspond to the initial position and initial `velocity' (tangent
vector). Since the norm-squared of the tangent vector is fixed by 
the non-extremality parameter $c$, one integration constant is 
determined by $c$. Equivalently, we can regard $c$ as being determined
by the integration constants of the scalar equations. 

Geodesics which lift to regular black string solutions need to 
satisfy specific boundary conditions. This will reduce the number
of independent integration constants. For static solutions,
irrespective of whether they are BPS or non-BPS,
we expect that solutions depend on $2n_V^{(5)} + 1$ 
integration constants, namely the $n_V^{(5)}+1$ magnetic charges
and the initial values of the $n_V^{(5)}$ physical scalar fields
at infinity. Due to the attractor mechanism, the values of the
scalars at the horizon are fixed in terms of the magnetic charges,
and therefore the number of integration constants in the second
order equations of motion is reduced by a factor of $\frac{1}{2}$. 
We will show that for certain models we can construct explicit
non-extremal solutions which depend on one additional parameter,
namely the non-extremality parameter $c$. The interpretation 
of the remaining integration constants will be discussed in 
Section 6, where we lift three-dimensional solutions to five
dimensions.

\subsection{The equation of motion for $\xi$}

The equation of motion for $\xi$ is the easiest to deal with. 
It reads $\ddot{\xi}=0$, which is solved by 
\[
\xi(\tau)=a\tau+b \;,
\] 
with two arbitrary constants $a,b$. However, there are
additional conditions which must be satisfied if
the three-dimensional solution lifts to a regular five-dimensional
black string. Transverse asymptotic flatness of (\ref{MetricAnsatzTrunc})
implies that $e^{2\sigma}$ and $e^\xi$
must independently approach unity for 
$\tau \rightarrow 0 \Leftrightarrow \rho
\rightarrow \infty$. For $\xi$ this implies that we must choose $b=0$ and
hence we have $\xi = a\tau$\,\footnote{The required asymptotics of $\sigma$ 
at infinity imposes conditions on the solutions for the other scalar
fields to which we will return later.}.
Next,
let us look at the near-horizon geometry $\tau\rightarrow\infty$. 
In this regime, the three-dimensional metric \eqref{3dMetric} behaves as
\[
ds^2_{(3)}\sim (2c)^4 e^{-4c\tau}d\tau^2+ (2c)^2 e^{-2c\tau}d\Omega^2_2 \quad \textrm{as }\,\tau\rightarrow\infty,
\]
so in this regime the full five-dimensional metric \eqref{MetricAnsatzTrunc} will look like
\begin{eqnarray}
ds^2_{(5)\textrm{hor}}&\sim & e^{2\sigma(\tau) +a\tau}
\left[-\epsilon_1e^{-a\tau}(dx^0)^2-\epsilon_2 e^{a\tau}(dx^4)^2\right]
\nonumber \\
&& +(2c)^2e^{-4\sigma(\tau)-2(a+c)\tau}\left[(2c)^2e^{-2c\tau}d\tau^2+d\Omega^2_2\right].
\end{eqnarray}
%where $\sigma_\infty$ is the (undetermined) horizon-value of the Kaluza-Klein scalar $\sigma$.

The horizon of the black string has topology $S^2 \times \mathbbm{R}$. 
In order to have a finite horizon size, both the metric coefficient
of the ``$S^2$-factor'' and of the ``$\mathbbm{R}$-factor'',
must be finite. Looking at the coefficient of the $d\Omega_{(2)}^2$-term,
we see that we must require
\[
2\sigma(\tau)=2\sigma_{\textrm{hor}}-(a+c)\tau \quad\textrm{as }\,\tau\rightarrow\infty,
\]
so that the line element becomes
\begin{eqnarray}
ds^2_{(5)\textrm{hor}}&=& e^{2\sigma_{\textrm{hor}}-c\tau}
\left[-\epsilon_1e^{-a\tau}(dx^0)^2-\epsilon_2 e^{a\tau}(dx^4)^2\right] \nonumber \\
&&
+(2c^2)e^{-4\sigma_{\textrm{hor}}}\left[(2c)^2e^{-2c\tau}d\tau^2+d\Omega^2_2\right].
\end{eqnarray}
Depending on whether we take $x^0$ or $x^4$ as the spatial 
coordinate along the string, we then need to take $a=-c$ 
or $a=c$ to have a finite coefficient for the $dx^0$-term or
$dx^4$-term, respectively. This condition can
be written in universal form as $a=\epsilon_1 c$, where
$\epsilon_1=-1=-\epsilon_2$ corresponds to space-time reduction,
while $\epsilon_1=1=-\epsilon_2$ corresponds to time-space reduction.
The solution for $\xi(\tau)$ is
\begin{equation}\label{xiSoln}
\xi(\tau)=\epsilon_1 c\tau.
\end{equation}
Note that the integration constants $a,b$ have been determined 
in terms of the non-extremality parameter $c$ by imposing boundary conditions. 
Thus the number of independent parameters has been reduced by 2.

\subsection{The equations of motion of $s_i$ }

We now move on to the equation of motion for the scalars $s_i$, which
were obtained by dualizing the three-dimensional gauge fields:
\[
\frac{d}{d\tau}\left(\hat{g}^{ij}(w)\dot{s}_j\right)=0.
\]
Integrating, we find
\begin{equation}\label{sSoln}
\dot{s}_i=\hat{g}_{ij}(w)\tilde{p}^j.
\end{equation}

In terms of the corresponding five-dimensional gauge fields we have
\begin{equation}
F^i_{\theta\varphi}=-\frac{1}{\sqrt{2}}\tilde{p}^i\sin\theta \;,
\end{equation}
and therefore we will refer to the parameters $\tilde{p}^i$ as the
magnetic charges carried by the string. While further integrating
(\ref{sSoln}) will introduce another  $n_V^{(5)}+1$ integration
constants, the metric is invariant under constant shifts of the
$s_i$, and therefore solutions where these integration constants
are chosen differently are related by isometries. From the five-dimensional
point of view such solutions are related by gauge transformations,
and therefore we will not count these integration constants as 
relevant parameters.

We note that, by substituting in (\ref{sSoln}),
the Hamiltonian constraint \eqref{HamConstraint} becomes
\begin{equation}\label{HamConstraint2}
\frac{3}{4}c^2-\hat{g}_{ij}(w)\left(\dot{w}^i\dot{w}^j -\tilde{p}^i \tilde{p}^j\right)=0.
\end{equation}
This will be useful in the following.

\subsection{The equation of motion of the $w^i$}
%%%

Finally, the equation of motion for the scalars $w^i$ reads, after making use of \eqref{sSoln},
\begin{equation}\label{wEqn}
\frac{d}{d\tau}\left(\hat{g}_{ij}(w)\dot{w}^j\right)
-\frac{1}{2}\left(\partial_i\hat{g}_{jk}(w)\right)\left(\dot{w}^j\dot{w}^k+\tilde{p}^j\tilde{p}^k\right)=0.
\end{equation}

Using the fact that $\hat{g}_{ij}$ is Hessian, this becomes
\begin{equation}
\label{wEqn2}
\hat{g}_{ij}(w)\ddot{w}^j+\frac{1}{2}\partial_i\hat{g}_{jk}\left(\dot{w}^j\dot{w}^k-\tilde{p}^j\tilde{p}^k\right)=0.
\end{equation}
Due to the explicit dependence on $\hat{g}_{ij}(w)$ and its derivatives,
it is difficult to solve this equation explicitly in a model-independent
way. We will proceed as in \cite{Mohaupt:2010fk} and find a class
of explicit solutions, which depending on the model might even 
be the general solution, and which at least
always contains a solution which recovers the standard RN
black string (with arbitrary charges but constant five-dimensional
scalar fields).  

To obtain this class of solutions we contract (\ref{wEqn2}) with $w^i$. 
Using the fact that $\hat{g}_{ij}$ is homogeneous of degree $-2$, we find
\begin{equation}
\hat{g}_{ij}(w)\left(w^i\ddot{w}^j-\dot{w}^i\dot{w}^j+\tilde{p}^i\tilde{p}^j\right)=0.
\end{equation}

Then, using \eqref{HamConstraint2} and the identity 
$\hat{g}_{ij}(w)w^i w^j=-\frac{3}{4}$, we arrive at the equation
\begin{equation}
\hat{g}_{ij}(w)w^i\left(\ddot{w}^j-c^2 w^j\right)=0.
\end{equation}

This equation still contains $\hat{g}_{ij}(w)$ but we can
obtain a class of universal, model-independent solutions
by setting $\ddot{w}^j-c^2 w^j=0$, which results 
in\footnote{The factors have been chosen for later convenience
with regard to taking the extremal limit.}:
\begin{equation}\label{wSoln}
w^i(\tau)=A^i\cosh(c\tau)+\frac{B^i}{c}\sinh(c\tau),
\end{equation}
where $A^i, B^i$ are constants. 
It remains of course to show that the 
full scalar equation of motion (\ref{wEqn2}) and the Hamiltonian
constraint \eqref{HamConstraint2} are solved. 

Substituting (\ref{wSoln}) into the Hamiltonian constraint
\eqref{HamConstraint2} gives
\[
\hat{g}_{ij} \left( c^2 A^i A^j - B^i B^j + \tilde{p}^i \tilde{p}^j \right)
=0 \;.
\]
Similarly, using $\ddot{w}^j-c^2 w^j=0$ the full scalar 
equation of motion \eqref{wEqn2} becomes
\[
\frac{1}{2} \partial_k \hat{g}_{ij} \left( c^2 w^i w^j 
- \dot{w}^i \dot{w}^j + \tilde{p}^i \tilde{p}^j \right) = 0 \;,
\]
and substituting in the explicit solution \eqref{wSoln} gives
\[
\frac{1}{2} \partial_k \hat{g}_{ij}  
\left( c^2 A^i A^j - B^i B^j + \tilde{p}^i \tilde{p}^j \right)=0 \;.
\]
Thus both remaining equations impose relations between the 
integration constants, which have to hold for each value
of $\tau$ separately, because the relations contain $\hat{g}_{ij}(w)$. 

At this point any further analysis depends on the form of $\hat{g}_{ij}(w)$.
For `diagonal models', 
where $\hat{g}_{ij}$ and $\partial_k \hat{g}_{ij}$ are diagonal
in $(i,j)$, we can solve both the Hamiltonian
constraint and the scalar equation of motion by imposing
the $n_V^{(5)} +1$ relations
\begin{equation}
\label{Constraint1}
c^2 (A^i)^2 - (B^i)^2 + (\tilde{p}^i)^2 = 0 \;.
\end{equation}
Thus we are left with $2n_V^{(5)} +2$ independent non-trivial 
integration constants for the scalar equations of motion.
Apart from fixing the integration constants for $\xi$ the
number of integration constants for $(w^i, s_i)$ were 
reduced by a factor of $\frac{1}{2}$, by discarding the irrelevant
initial values of $s_i$ and by imposing \eqref{Constraint1}. 
By later investigation of the resulting five-dimensional black 
string solutions we will see that \eqref{Constraint1} can 
be viewed as a deformed version of the black hole attractor
mechanism, which determines half of the integration constants
of the five-dimensional scalars in terms of the magnetic 
charges. The diagonal models include the $ST^2$ and $STU$ models
of five-dimensional supergravity and $STU$-like models
in non-supersymmetric theories constructed using generalized special
real geometry.

For non-diagonal models the ansatz \eqref{wSoln} only yields
solutions with a reduced number of integration constants. 
In the most generic case, where $\hat{g}_{ij}$ and its derivatives (when
evaluated on the solution) do not allow a simultaneous 
block decomposition, the only model-independent 
way to make the ansatz \eqref{wSoln} work is to impose the
stronger condition 
\[
c^2 A^i A^j - B^i B^j + \tilde{p}^i \tilde{p}^j = 0 \;.
\]
The additional off-diagonal relations can still be 
solved by imposing 
\[
\frac{A^i}{A^j} = \frac{B^i}{B^j} = \frac{\tilde{p}^i}{\tilde{p}^j},
\]
but this has the effect that the ratios 
$\frac{w^i(\tau)}{w^j(\tau)}$ are constant,
so that all scalar fields $w^i(\tau)$ are proportional to one 
another. 
From the formulae given below it will be clear that in 
this case the five-dimensional metric is just the one of the
standard RN-type black string. The physical five-dimensional scalars,
which can be chosen to be parametrized by 
the $n_V^{(5)}$ independent ratios of the
fields $w^i$, are constant for this universal solution.

In between these extremes are models where $\hat{g}_{ij}$ and
$\partial_k \hat{g}_{ij}$ admit a 
simultaneous decomposition into $k$ 
different blocks ($k=1$ is the most generic indecomposable
case discussed in the previous paragraph). For such models
we obtain $k$ sets of non-proportional scalars 
$w^i$. Thus for $k>1$ the solutions will admit $k-1 >0$ 
independent non-constant five-dimensional scalars, 
and the five-dimensional metric will be  
different from the standard RN-type black string metric.

Furthermore, the ansatz \eqref{wSoln} might still yield
non-trivial solutions for indecomposable scalar metrics,
if one can restrict the solution to a 
totally geodesic submanifold of $S$, on which the
metric becomes block-decomposable. Examples of this phenomenon
were observed in \cite{Mohaupt:2011aa}.

\section{Non-extremal black string solutions}\label{Solns}

We now proceed to investigate the black string solutions
obtained by the ansatz (\ref{wSoln}). To prepare for this
we rewrite \eqref{wSoln}  in terms of the new radial coordinate $\rho$ defined in \eqref{IsoCoords}:
\begin{equation}\label{wSoln2}
w^i(\rho)=\left(A^i+\frac{p^i}{\rho}\right)W^{-\frac{1}{2}}:=\mathcal{H}^i(\rho)W^{-\frac{1}{2}}.
\end{equation}

Here we have used the definition of the function $W(\rho)$ given in \eqref{NonExtrHarm}, and introduced $p^i:=B^i-cA^i$. At this point it is convenient also to introduce the quantity $\bar{p}^i:=p^i+2cA^i$. We will see later on that $p^i$ and $\bar{p}^i$ are related to the values of the scalar fields $h^i(\rho)$ at the inner and outer horizons respectively.
We also note for later reference that 
in terms of the charges $p^i, \bar{p}^i, \tilde{p}^i$ the 
Hamiltonian constraint takes the form
\[
\hat{g}_{ij} (\tilde{p}^i \tilde{p}^j - p^i \bar{p}^j) =0 \;.
\]

%The solution of the reduced three-dimensional theory 
%found in the previous section is essentially
%given in terms of the three-dimensional scalars $w^i$ 
%by the expression \eqref{wSoln} for $w^i(\tau)$, or, 
%equivalently, \eqref{wSoln2} for $w^i(\rho)$. To obtain
%the corresponding five-dimensional black string solution,
We now express the solution in terms of five-dimensional
quantities. 
Using \eqref{TruncFields} and the hypersurface constraint \eqref{Hypersurface}, we see that
\[
e^{\xi+2\sigma}=H(w)^{-\frac{1}{3}}=H(\mathcal{H})^{-\frac{1}{3}}W^{\frac{1}{2}},
\]
so, using also \eqref{xiSoln} and \eqref{NonExtrHarm}, the five-dimensional metric \eqref{MetricAnsatzTrunc} becomes
\begin{equation}\label{BlackStringLE2}
ds^2_{(5)}=H(\mathcal{H})^{-\frac{1}{3}}\left(- W dt^2 + dy^2\right)
+H(\mathcal{H})^{\frac{2}{3}}\left(\frac{d\rho^2}{W}+\rho^2 d\Omega_2^2\right),
\end{equation}
where $\lbrace t,y\rbrace$ are the time-like and space-like directions corresponding to the worldvolume of the string. We note that this form of the solution is \textit{independent} of which order (space-then-time or time-then-space) we perform the reduction. The metric \eqref{BlackStringLE2} is a 
generalization of the standard RN black string metric, where the single
harmonic function $\mathcal{H}$ has been replaced by the function 
$(H(\mathcal{H}^i))^{1/3}$, which depends on $n_V^{(5)}+1$
harmonic functions $\mathcal{H}^i(\rho)$. The standard RN-type string is recovered
when all these harmonic functions are proportional to one another.

The (constrained) five-dimensional scalar fields are given by
\begin{equation}
h^i(\rho)=H(\mathcal{H})^{-\frac{1}{3}}\mathcal{H}^i(\rho).
\end{equation}

Transverse asymptotic flatness of the metric implies that
$H(\mathcal{H}) \rightarrow 1$ for $\rho \rightarrow \infty$.
Therefore the constant term $A^i$ in the harmonic function $\mathcal{H}^i$ 
specifies the value of the scalar $h^i$ at transverse infinity,
$A^i = h^i_\infty$, and 
\[
\mathcal{H}^i(\rho)=h^i_\infty+\frac{p^i}{\rho}.
\]
The condition of transverse asymptotic flatness
$H \rightarrow 1$ can be written as
$H(h_\infty) =1$ by taking the limit. This imposes
one relation between the $n_V^{(5)}+1$ integration constants
$h^i_\infty$. Obviously, this condition is precisely the 
hypersurface constraint and takes into account the fact that there are
only $n_V^{(5)}$ independent five-dimensional scalars for which
we can impose boundary values at infinity. One convenient way
to parametrize the independent five-dimensional scalars is
to use $n_V^{(5)}$ independent ratios, for 
example  $\phi^x = \frac{h^x}{h^0} = \frac{w^x}{w^0}$ 
\cite{Mohaupt:2010fk,Mohaupt:2012tu}.

To interpret the integration constants $p^i$ (equivalently $B^i$) we
consider the limits $\rho\rightarrow 2c$ and $\rho\rightarrow 0$, which correspond to the outer and inner horizons respectively. We see that, in these cases, the scalars $h^i(\rho)$ satisfy
\[
h^i\xrightarrow[\rho\rightarrow 2c]{}\left(H(\bar{p})(2c)^{-3}\right)^{-\frac{1}{3}}\frac{\bar{p}^i}{2c}=H(\bar{p})^{-\frac{1}{3}}\bar{p}^i,
\]
and
\[
h^i\xrightarrow[\rho\rightarrow 0]{}\left(H(p)\rho^{-3}\right)^{-\frac{1}{3}}\frac{p^i}{\rho}=H(p)^{-\frac{1}{3}}p^i.
\]
Here we use $p^i := B^i - cA^i$ and $\bar{p}^i := p^i + 2cA^i$. 

This is the same ``dressed attractor behaviour'' as noted in 
\cite{Mohaupt:2010fk} for five-dimensional black holes
and motivates calling $\bar{p}^i$ and $p^i$ the outer and 
inner ``horizon charges'' respectively. It remains to clarify how these
``horizon charges'' are related to the physical magnetic charges 
$\tilde{p}^i$. To do this recall that the magnetic charges are (the non-trivial
half of) the integration constants of the scalars $s_i$ and appear
in the five-dimensional gauge fields as
\begin{equation}\label{FieldStrength}
F^i=-\frac{1}{\sqrt{2}}\tilde{p}^i\sin\theta\,d\theta\wedge d\varphi.
\end{equation}

%%%

As observed at the end of the previous section, the 
Hamiltonian constraint takes the form
\begin{equation}\label{HamConstraint3}
\hat{g}_{ij}(w)\left(\tilde{p}^i\tilde{p}^j-p^i\bar{p}^j\right)=0 \;.
\end{equation}
For diagonal models we solve this by imposing
\begin{equation}
\label{Constraint2}
(\tilde{p}^i)^2 - p^i \bar{p}^i = 0 \;,
\end{equation}
which is \eqref{Constraint1} expressed in terms of 
$p^i$ and $\bar{p}^i=p^i + 2ch^i_\infty$. This can 
be used to express the horizon charges $p^i$ (and, hence,
$\bar{p}^i$) in terms of $\tilde{p}^i$, $h^i_\infty$ and $c$:
\[
p^i=-ch^i_\infty\pm\sqrt{(\tilde{p}^i)^2+c^2(h^i_\infty)^2}.
\]
The sign is to be  chosen such that the metric is regular outside the 
horizon\footnote{We will come back to questions of regularity at
the end of Section 6.}.
We have now identified the number and interpretation of the 
independent integration constants for solutions \eqref{wSoln}
for diagonal models. There are $2n_V^{(5)} + 2$ independent
integration constants, namely $n_V^{(5)} +1$ magnetic charges
$\tilde{p}^i$, the $n_V^{(5)}+1$ constants $h^i_\infty$ which are
subject to one constraint and encode the asymptotic values
of the $n_V^{(5)}$ five-dimensional scalars at infinity, and
the non-extremality parameter $c$. 

A priori, one might have expected $n_V^{(5)}$ further integration
constants, corresponding to the initial velocities of the 
five-dimensional scalars at infinity, or, equivalently, 
their values at the outer or inner horizon. However, these
values are determined by the condition which generalizes the
attractor mechanism known from extremal black holes. While
we do not have proper fixed point behaviour, i.e. the values
of the scalars at the horizons are not determined exclusively
by the charges, but also depend on their values at infinity,
it is still true that there are no independent integration 
constants related to the horizon values, but rather they are 
determined by other data.
This  suggests that the solution can be obtained from a reduction of the
scalar field equations to first order form, similar to BPS
equations. As discussed in \cite{Mohaupt:2012tu} for the similar
case of five-dimensional black holes, the deformed attractor 
mechanism guarantees that the physical scalar fields take
finite values on the horizon.

%%%here

%{\bf TMM: Explain relation of $Z$ to magnetic central charge } \\[2ex]

Let's now turn our attention to some further properties of the solution.

%\textbf{PD: Add NH geometry for the outer horizon, as well as tension, central charge for non-extremal solutions.}

In order to explore the geometry near the outer horizon, we introduce the variable $u^2=\rho-2c$, and look at the region $u^2\approx 0$. Then \eqref{BlackStringLE2} becomes
\begin{equation}
ds^2_{(5)}=\frac{2c}{H(\bar{p})^\frac{1}{3}}dy^2 +H(\bar{p})^{\frac{2}{3}}d\Omega^2_2 +\frac{2H(\bar{p})^{\frac{2}{3}}}{c}\left(du^2-\frac{c}{2H(\bar{p})}u^2dt^2\right).
\end{equation}

Introducing $v^2=\rho$ and concentrating on the region $v^2\approx 0$, 
we find that the metric near the inner horizon takes the form
\begin{equation}\label{NearInnerHorizon}
ds^2_{(5)}=\frac{2c}{H(p)^\frac{1}{3}}dt^2 +H(p)^{\frac{2}{3}}d\Omega^2_2 +2\frac{H(p)^{\frac{2}{3}}}{c}\left(-du^2+\frac{c}{2H(p)}u^2dy^2\right).
\end{equation}

%\textbf{PD: The $Z$ here shouldn't be confused with the central charge $\mathcal{Z}$, which depends on the asymptotic data.}

In both cases, the first two factors give an $\mathbb{R}\times S^2$, with the size of the $S^2$ determined by the horizon charges $p^i,\bar{p}^i$, whilst the rest of the metric takes the form of a two-dimensional Rindler space.
%
%\textbf{PD: Comments? Interpolating between vacua. Outer horizon?}

From these expressions we can read off that the entropy of the inner and outer horizons are given, respectively, by
\[
S_-=\pi H(p)^{\frac{2}{3}}, \hspace{5mm} S_+=\pi H(\bar{p})^{\frac{2}{3}},
\]
whilst the temperatures associated to each horizon are
\[
T_-=\frac{\sqrt{2c}}{4\pi}H(p)^{-\frac{1}{2}}, \hspace{5mm} T_+=\frac{\sqrt{2c}}{4\pi}H(\bar{p})^{-\frac{1}{2}},
\]
which vanish as expected in the extremal limit. The combination

\[
T_{\pm}S_{\pm}^{\frac{3}{4}}=\frac{\sqrt{2c}}{4}\pi^{-\frac{1}{4}},
\]
depends only on the non-extremality parameter.

%\textbf{PD: Comments.}

The tension of the solution is
\[
\mathcal{T}=\frac{1}{2}c_{ijk}h^i_\infty h^j_\infty \bar{p}^k,
\]
where we are using the normalization of \cite{Martin:2012bi}.

%{\bf TMM: Add Definitions/References. Action on Paul.}

We conclude our discussion by pointing out that in order 
to obtain regular black string solutions one might need
to impose further conditions in addition to the restrictions
that guarantee asymptotic flatness and a regular solution
on the horizon. 
The line element is modified compared to the standard 
RN black string by replacing the single harmonic function $\mathcal{H}(\rho)$
by $(H(\mathcal{H}^i))^{1/3}$, which is a rational function 
of several harmonic functions. Therefore it may happen that,
for some choices of integration
constants, $(H(\mathcal{H}^i))^{1/3}$ takes 
the values zero or infinity at finite $\rho > 2c$, generically resulting
in a naked singularity even if the behaviour at $\rho \rightarrow
\infty$ and $\rho = 2c$ is regular. This phenomenon was studied
for five-dimensional BPS black holes and five-dimensional domain 
walls in \cite{Kallosh:2000rn} and \cite{Mayer:2003zk}. It was
observed in particular that naked singularities can occur even though
the scalar fields take finite values within the scalar manifold along
the whole solution. For M-theory compactifications on 
Calabi-Yau threefolds naked singularities cannot occur for domain 
walls and BPS black holes as long as the scalar fields take
values within the extended K\"ahler cone, which is the modified scalar manifold
relevant for M-theory \cite{Mayer:2003zk}. However, apart from 
this there are no model-independent results we are aware of. For the
case at hand, we should therefore add the condition that the
integration constants $(h^i_\infty, p^i)$ have to be chosen such 
that $(H({\cal H}^i))^{1/3}$ does not have zeros or infinities for $\rho>2c$, 
and, if we want to continue the solution to the inner horizon,
for $\rho >0$. The existence of such solutions is guaranteed because
the standard RN black string is always contained in our class of
solutions. Sufficiently small deformations away from this solution
will not introduce zero or infinities for $(H({\cal H}^i))^{1/3}$ and
therefore give rise to regular solutions with non-constant scalar fields.
However, it cannot be excluded without model-by-model investigation
that large deformations away from the RN black string lead to singular
solutions.

\section{Extremal black strings}

%We turn now to the extremal $c\rightarrow 0$ limit of the black string solutions constructed in Section \ref{Solns}. The only change in the metric \eqref{BlackStringLE2} is that the function $W$ will become unity in this limit, as can be seen from the expression \eqref{NonExtrHarm}.
%One can of course construct more general multi-centred solutions in the extremal case, but this will simply correspond to a change in the harmonic functions $H^i(\rho)$ and won't affect the functional form of \eqref{BlackStringLE2}.

Extremal solutions can be obtained by either taking the limit $c\rightarrow
0$ of non-extremal solutions, or by directly solving the equations of
motion for $c=0$. To illustrate the drastic simplification occurring
in this limit, observe that for $c=0$ the Hamiltonian constraint 
simplifies to
\[
\hat{g}_{ij} ( \dot{w}^i \dot{w}^j - \tilde{p}^i \tilde{p}^j) = 0 \;,
\]
which can be solved, for any $\hat{g}_{ij}$, by 
\[
\dot{w}^i = p^i = \pm \tilde{p}^i \;,
\]
so that the solution of (\ref{wEqn})  is simply
\[
w^i = A^i + p^i \tau = h^i_\infty + \frac{p^i}{\rho} = \mathcal{H}^i(\rho) \;.
\]
Since $W=1$ there is only one horizon, and the horizon charges 
are equal to one another and, up to an overall sign, equal
to the magnetic charges: $p^i=\bar{p}^i = \pm \tilde{p}^i$. 
Further simplified relations include $\rho = \frac{1}{\tau}$ and
$\xi=0$.

%%%%%%%%%%%%%

At the horizon, the values of the scalars are determined by the  
charges $p^i = \pm \tilde{p}^i$:
\[
h^i \rightarrow H(p)^{-1/3} p^i \;,\;\;\;
\mbox{for} \;\;\;\rho \rightarrow 0 \;.
\] 
This is the attractor mechanism for BPS solutions. 

The ADM tension carried by an extremal string is
\[
\mathcal{T} =\frac{1}{2} c_{ijk} h^i_\infty h^j_{\infty} p^k,
\]
where $p^k$ are the parameters appearing in the solution for the
scalar fields, while the magnetic central charge is 
\cite{Antoniadis:1995vz,Chou:1997ba}
\[
\mathcal{Z}_m = h_i(\infty) \tilde{p}^i = c_{ijk} h^i_\infty h^j_{\infty}
\tilde{p}^k \; ,
\]
where $\tilde{p}^k$ are the magnetic charges.

Solutions where
$p^i=\pm \tilde{p}^i$ saturate the supersymmetric mass bound
\[
\mathcal{T} \geq \frac{1}{2}|\mathcal{Z}_m|,
\]
and are therefore BPS solutions. Promoting $\mathcal{Z}_m$ to a space-time field 
by setting $\mathcal{Z}_m = h_i \tilde{p}^i$, where $h_i = c_{ijk} h^j h^k$, 
one finds 
\[
\mathcal{Z}_m \rightarrow \frac{1}{\rho^2} c_{ijk} \tilde{p}^i \tilde{p}^j \tilde{p}^k
= \frac{1}{\rho^2} H(\tilde{p}) ,
\]
so that the attractor mechanism takes the form
\[
\mathcal{Z}_m h^i \rightarrow p^i \;\;\;\mbox{for}\;\;\;\rho \rightarrow 0 \;.
\]

In general, only $p^i = \pm \tilde{p}^i$ is guaranteed to 
give a solution of the Hamiltonian constraint and of the
field equations. But further solutions arise whenever the
scalar metric $\hat{g}_{ij}$ (when evaluated on the solution) 
admits a non-trivial `charge rotation matrix.' This observation was 
made in the context of first order flow equations
\cite{Ceresole:2007wx,LopesCardoso:2007ky}, but can be
applied to the second order formalism used here as previously 
in \cite{Mohaupt:2009iq,Mohaupt:2011aa}. A charge rotation
matrix is a constant matrix
which relates the horizon charges $p^i$ and the magnetic
charges $\tilde{p}^i$ by
\[
\tilde{p}^i = R^i_j p^j,
\]
and satisfies $\hat{g}_{ij} R^i_k R^j_l = \hat{g}_{kl}$
so that the Hamiltonian constraint (and the full field
equations (\ref{wEqn})) is solved. Such solutions are extremal, i.e.
have $c=0$ and a single horizon located at $\rho=\frac{1}{\tau} = 0$,
but they are not BPS because $\mathcal{T}\not=\frac{1}{2}|\mathcal{Z}_m|$.

For extremal solutions the assumption of three-dimensional 
spherical symmetry is not necessary, and by relaxing it we can
obtain multi-centred solutions. First note that for extremal 
solutions we have $\xi=0$, so that the scalar $\xi$ can
already be truncated out at the level of the action \eqref{BSAction2}.
In this case the target space of the three-dimensional theory
reduces to the para-K\"ahler manifold $N=T^*M$. We can then proceed 
essentially as in \cite{Mohaupt:2009iq}, with the minor modification
that there the para-K\"ahler manifold was $TM$, the tangent bundle
of a Hessian manifold $M$, rather than the cotangent bundle. 
Imposing the ``extremal instanton ansatz''
\[
\partial_\mu w^i = R^i_j \hat{g}^{jk}(w) \partial_\mu s_k,
\]
the equations of motion for $w^i$ reduce to 
\[
\Delta w^i = 0,
\]
where $\Delta$ is the flat three-dimensional Laplacian. 
Taking the solutions to be multi-centred Harmonic functions,
\[
w^i(\vec{x}) = \mathcal{H}^i(\vec{x}) \equiv 
h^i_\infty+\sum_{n}\frac{p_n^i}{|\vec{x}-\vec{x}_n|},
\]
where $\vec{x} = (x^\mu) = (x^1, x^2, x^3)$, we obtain static
multi-centred black string solutions with horizons located
at $\vec{x}_n$ in transverse space. The spherically symmetric 
solutions are recovered by restricting to solutions with one centre. 
The near horizon asymptotics of each centre is the same as for
the corresponding single-centred solutions. We do not give any
further details but refer to the analogous case of black holes
which was analysed in detail in \cite{Mohaupt:2009iq}.

\subsection{Example: $ST^2$ model}

We now choose a particular Hesse potential \eqref{Hypersurface} describing the one-dimensional special real manifold $h^0(h^1)^2=1$. Since BPS and non-BPS
black string solutions for this model have already been discussed
in \cite{Martin:2012bi}, we keep the presentation brief, with the
main purpose of comparing our formalism to the FGK formalism used
there. 
In order for the hypersurface $h^0(h^1)^2=1$
to be well-defined we must take $h^0>0$. There are then two disjoint patches in which $h^1$ can take values, namely $\lbrace h^1>0\rbrace$ and $\lbrace h^1<0\rbrace$. Working out the associated metric $\hat{g}_{ij}$, we find
\[
\hat{g}_{ij}=\frac{1}{4} \left(\begin{array}{cc}
(h^1)^4 & 0 \\
0 & 2h^0 
\end{array}\right).
\]

It turns out that there are 4 possible ``R-matrices'' satisfying $R^T\hat{g}R=\hat{g}$, namely $R=\pm R_{(\sigma)}$, where
\[
R_{(\sigma)}=\left(\begin{array}{cc}
1 & 0 \\
0 & \sigma
\end{array}\right),
\]
and $\sigma=\pm 1$.

%{\bf TMM: new text inserted}

For this model the ADM tension $\mathcal{T}$ is given by
\[
6\mathcal{T} = (h^1_\infty)^2 p^0 + 2 h^0_\infty h^1_\infty p^1 ,
\]
where $p^0, p^1$ are the horizon charges, while the 
magnetic central charge $\mathcal{Z}_m$ is given by
\[
3 \mathcal{Z}_m = (h^1_\infty)^2 \tilde{p}^0 + 2 h^0_\infty h^1_\infty \tilde{p}^1 , 
\]
where $\tilde{p}^0$, $\tilde{p}^1$ are the magnetic charges. Let us
discuss the range of values that the parameters $h^i_\infty, p^i, \tilde{p}^i$
can take. The magnetic charges $\tilde{p}^i$ can 
independently be positive or negative. In contrast the parameters
$h^i_\infty, p^i$ are restricted by the fact that the scalar fields
\[
h^i \simeq \mathcal{H}^i = h^i_\infty + \frac{p^i}{\rho},
\]
must take values inside the scalar manifold for $\infty > \rho >0$.
For definiteness, consider the connected component
$\{ h^0>0, h^1>0 \}$. Then we must impose that all four parameters
are positive: $h^0_\infty>0$, $h^1_\infty>0$, $p^0>0$, $p^1>0$.
This implies immediately that solutions where $R=\pm R_{(1)}$
saturate the BPS bound $\mathcal{T} = \frac{1}{2} |\mathcal{Z}_m|$ while for $R= \pm R_{(-1)}$
we have $\mathcal{T} > \frac{1}{2} |\mathcal{Z}_m|$. Thus the solutions generated by a 
non-trivial charge rotation matrix are non-BPS. We note that
on the component $\{h^0>0, h^1 > 0\}$ of the scalar manifold
BPS solutions have magnetic charges with the same sign
(i.e. both positive or both negative) while non-BPS solutions
have magnetic charges with opposite signs. 

One can also consider the second connected
component $\{h^0>0, h^1 < 0 \}$. On this component BPS solutions
have opposite signs of the magnetic charge while non-BPS solutions 
have magnetic charges with the same sign.
Our results for the $ST^2$ model are consistent
with those of \cite{Martin:2012bi}. One distinct feature
of our formalism, which we view as an advantage, is that we can
perform the whole analysis using the homogeneous coordinates
$(h^0,h^1)$, without making a choice for a physical scalar parametrizing
the hypersurfaces. The FGK formalism used in \cite{Martin:2012bi}
requires such a choice, in order to minimize the effective potential, 
describe attractor behaviour, and to identify the different branches
corresponding to BPS and non-BPS solutions. In contrast we can obtain
the same information more easily working in homogeneous coordinates.

\subsection{M-theory compactifications on Calabi-Yau threefolds}

We remark that there is an important class of models where
the domain of the scalar fields can be chosen of the
form $\{h^i > 0 \}$, namely compactifications of M-theory
on toric Calabi-Yau threefolds. In this case the scalar manifold
is the hypersurface of the K\"ahler cone of the Calabi-Yau
manifold obtained by fixing the volume. For toric Calabi-Yau
threefolds the K\"ahler cone is a `strongly convex finite
polyhedral cone', which admits a parametrization of the above
form. We refer to \cite{Mayer:2003zk} and references therein 
for details. 
In this parametrization, all charges will be either
positive or negative for BPS solutions, while non-BPS solutions,
if they exist, will have a mixture of positive and negative 
charges. In general the metric will not have a block decomposition,
so that we cannot guarantee the existence a non-trivial charge rotation matrix 
and, hence, of explicit non-BPS solutions.

\subsection{Example: $STU$ model}

As a final illustration of our method for constructing BPS and non-BPS extremal solutions, we consider the case of the $STU$ model, which has Hesse potential $H(h)=h^0h^1h^2$. Again, we keep the discussion brief as the extremal BPS and non-BPS solutions to this
model have been discussed before using the FGK formalism 
in \cite{Meessen:2012su}.

The equation $h^0h^1h^2=1$ 
defines a two-dimensional projective special real manifold, which consists of four disjoint patches depending on the signs of (say) $h^0$ and $h^1$.
The metric $\hat{g}_{ij}$ is

\[
\hat{g}_{ij}=\frac{1}{4} \left(\begin{array}{ccc}
(h^1h^2)^2 & 0 & 0 \\
0 & (h^0h^2)^2 & 0 \\
0 & 0 & (h^0h^1)^2
\end{array}\right).
\]

The eight possible charge rotation matrices satisfying $R^T\hat{g}R=\hat{g}$ in this case are given by $R=\pm R_{(\sigma,\tau)}$, where
\[
R_{(\sigma,\tau)}=\left(\begin{array}{ccc}
1 & 0 & 0 \\
0 & \sigma & 0 \\
0 & 0 & \tau
\end{array}\right),
\]
and $\sigma$ and $\tau$ can each take the values $\pm 1$.

The ADM tension $\mathcal{T}$ and magnetic central charge $\mathcal{Z}_m$ are given, respectively, by
\[
6\mathcal{T}=h^0_\infty h^1_\infty p^2+h^0_\infty h^2_\infty p^1 +h^1_\infty h^2_\infty p^0,
\]
and
\[
3\mathcal{Z}_m=h^0_\infty h^1_\infty \tilde{p}^2+h^0_\infty h^2_\infty \tilde{p}^1 +h^1_\infty h^2_\infty \tilde{p}^0.
\]

Taking, for concreteness, the patch $\lbrace h^0>0, h^1>0\rbrace$, we find again that solutions with $R=\pm R_{(1,1)}$ saturate the BPS bound $\mathcal{T}=\frac{1}{2}|\mathcal{Z}_m|$, whilst for the six other choices of $R$ we have $\mathcal{T}>\frac{1}{2}|\mathcal{Z}_m|$.

We note that for any diagonal model one can always find an $R$-matrix
which flips the sign of any of the charges $\tilde{p}^i$. Thus for
diagonal models we cannot only find explicit non-extremal solutions,
but also explicit extremal solutions with any choice of signs for the
charges. Moreover, this does not only apply to supergravity models,
but also to non-supersymmetric models with couplings determined
by generalized special real geometry, as we will see in 
Section \ref{Section_GSG}.

\section{Small black holes}

The method of dimensional oxidation employed in Section \ref{Solns} to obtain black string solutions of the original five-dimensional action \eqref{5dAction} can also be used to generate a class of four-dimensional black hole solutions to the spacelike reduction of \eqref{5dAction}.

In particular, we can take the three-dimensional solutions constructed in Section \ref{Instantons} and lift them over a single timelike direction, thereby obtaining a solitonic solution to a four-dimensional action.
The line element we obtain is
\begin{equation}\label{NonExtSBH}
ds^2_{(4)}=-H(\mathcal{H})^{-\frac{1}{2}}Wdt^2+H(\mathcal{H})^{\frac{1}{2}}\left(\frac{d\rho^2}{W}+\rho^2 d\Omega^2_2\right),
\end{equation}
which corresponds to a black hole solution having an inner horizon at $\rho=0$ and an outer horizon at $\rho=2c$. The area of the outer horizon is
\[
A_+=4\pi\sqrt{2c}\,H(\bar{p})^{\frac{1}{2}},
\]
whereas the area of the inner horizon vanishes. In the extremal limit
$c\rightarrow 0$ the outer horizon shrinks to zero size, so we
are left with what has been dubbed a `small' black hole. In the context
of string theory, black hole solutions are modified by higher
derivative corrections to the effective action 
\cite{LopesCardoso:1998wt,LopesCardoso:2000qm}, which has the
effect that small (extremal) black holes obtain a finite
horizon \cite{Dabholkar:2004dq}. The non-extremal 
black holes solutions obtained above are non-extremal deformations
of such small black holes. Our solutions show that while non-extremality
makes the outer horizon of small black holes finite, the inner horizon
still remains singular in the absence of higher derivative corrections.

%%%here

\section{Generalized special geometry \label{Section_GSG}}

As emphasized in Section \ref{dimred}, the formalism we have used above in constructing non-extremal black string solutions depends on
$H$ being a homogeneous function, but not on its degree or polynomial nature.
In the previous section we took $H$ to be of degree three for concreteness.
Let us now see what changes if we take $H$ to have a different degree.

To start with, the five-dimensional 
vector kinetic coupling $a_{ij}$ is still given by (\ref{aij}) 
in terms of a homogeneous function $H$ and is thus homogeneous
of degree $-2$. Moreover, the physical
scalar manifold is still given by the level set $\{H=1\}$.
However the constant factor $c_{ijk}$ in front of the Chern-Simons
term is no longer related to the function $H$. In supergravity theories 
supersymmetry relates the Chern-Simons term to other terms in the
action and forces the coefficient to be given by the third derivatives
of $H$. Gauge symmetry (up to a surface term) then implies that 
the third derivatives of $H$ must be constant, thus forcing 
$H$ to be a homogeneous degree three polynomial. If 
we change the degree of homogeneity and thus give up
supersymmetry, gauge symmetry still forces $c_{ijk}$ to be constant,
but it is no longer encoded by the function $H$ and becomes an
independent set of parameters. As far as purely magnetic black string solutions
are concerned (or purely electric black hole solutions) these 
parameters are however irrelevant, because the Chern-Simons term
does not contribute to purely magnetic (or purely electric) solutions.
Thus for this class of solutions the only input needed is the function
$H$, which we take to be homogeneous of degree $n$.

The dimensional reduction proceeds as before with some changes of
numerical factors in some formulae.
The explicit expression for $\hat{g}_{ij}(y)$ given in \eqref{gMetric} will be modified, although it will still take the form
\[
\hat{g}_{ij}(y)=-\frac{1}{4}\partial_{ij}^2\log H(y),
\]
with $H$ a homogeneous function. The expression \eqref{BSAction2} for the
reduced action remains valid, and since $(M,\hat{g}_{ij})$ is a Hessian
manifold, the target space $S$ of the reduced theory 
is still the product of the para-K\"ahler manifold
$N\simeq T^*M$ with a one-dimensional factor parametrized by $\xi$.
While this follows from known results
\cite{2008arXiv0811.1658A,Cortes:2009cs,Mohaupt:2009iq}, 
we give a short self-contained 
proof in Appendix \ref{AppHessianPK}.

We can then follow through the construction of non-extremal black string solutions as in Sections \ref{Instantons} and \ref{Solns} above. The main difference is in the form of the line element \eqref{BlackStringLE2}, which becomes
\begin{equation}\label{BlackStringLEgen}
ds^2_{(5)}=H(\mathcal{H})^{-\frac{1}{n}}\left(- W dt^2 + dy^2\right)
+H(\mathcal{H})^{\frac{2}{n}}\left(\frac{d\rho^2}{W}+\rho^2 d\Omega_2^2\right).
\end{equation}

For example, one could consider the `$STU$-like' models 
introduced  in \cite{Mohaupt:2009iq},
which have the Hesse potential
\[
H(h)=h^1\ldots h^n.
\]
In this case the line element \eqref{BlackStringLEgen} takes the form
\begin{equation}
ds^2_{(5)}=\frac{1}{(\mathcal{H}^1\ldots \mathcal{H}^n)^\frac{1}{n}}\left(- W dt^2 + dy^2\right)
+(\mathcal{H}^1\ldots \mathcal{H}^n)^{\frac{2}{n}}\left(\frac{d\rho^2}{W}+\rho^2 d\Omega_2^2\right),
\end{equation}
where each of the $\mathcal{H}^i(\rho)$ are harmonic functions. The scalar fields $h^i(\rho)$ are given by
\begin{equation}
h^i(\rho)=\frac{\mathcal{H}^i(\rho)}{(\mathcal{H}^1\ldots \mathcal{H}^n)^{\frac{1}{n}}}.
\end{equation}

For the case where all of the $\mathcal{H}^i\propto \mathcal{H}$ are proportional to one another, we find that the scalar fields $h^i(\rho)$ take constant values, and the line element collapses to that of the RN black string \eqref{5dRNBS}.

As in the supersymmetric case, all models admit generic `BPS-type'
extremal solutions where $p^i=\pm \tilde{p}^i$, while further 
explicit solutions can be found whenever a charge rotation matrix
exists. All such extremal solutions admit non-spherical, multi-centred
versions. For STU-like models the discussion given for $ST^2$ and $STU$
model can be adapted. For these models there exist charge rotation 
matrices which allow one to find explicit solutions for any choice of
signs for the charges. 

Since all this is completely analogous to the case of five-dimensional
black hole solutions discussed in \cite{Mohaupt:2009iq}, we refrain from
giving more details or working through explicit examples, 
but instead discuss the relation between 
geodesics in the manifold $S=N \times \mathbbm{R}$ and five-dimensional
black string solutions from a general geometrical point of view. 
%The relevant mathematical results have been proved in 
%\cite{Cortes:2003zd,Cortes:2005uq,Cortes:2009cs}.
To start, let us remember that while non-extremal solutions 
correspond to space-like geodesics in $N \times \mathbbm{R}$, 
extremal solutions correspond to null geodesics in $N$. 
If we do not assume the existence of a charge rotation matrix,
we can still always find explicit extremal solutions which satisfy the same 
relation 
\begin{equation}
\label{BPS-type}
\partial_\mu {w}^i = \pm \hat{g}^{ij} \partial_\mu {s}_j,
\end{equation}
as BPS solutions in supersymmetric theories. We refer to such 
solutions as BPS-type solutions. Using the information about
the para-K\"ahler geometry of the manifold $N$ collected in
Appendix \ref{AppHessianPK}, we obtain a geometric characterisation
of BPS-type solutions, which does not make use of supersymmetry and
applies to BPS-type solutions of non-supersymmetric theories as well.
Comparing (\ref{BPS-type}) to formula (\ref{IsotropicFrame})
in Appendix \ref{AppHessianPK} it is manifest that the BPS-type
solutions evolve along the `eigendirections' (eigendistributions)
of the para-complex structure of $N$. As explained in 
\cite{Mohaupt:2009iq} the integral submanifolds tangent to these
eigendirections are not only isotropic and totally geodesic
(hence solving the equations of motion) but even flat,
which explains why the solution can be written in terms of harmonic
functions. 

If the metric admits a non-trivial charge rotation 
matrix we can explicitly construct
further extremal solutions, which satisfy
\[
\partial_\mu w^i = R^i_j \hat{g}^{jk} \partial_\mu s_k \;,
\]
with $R^i_j\not=\pm \delta^i_j$. 
For supersymmetric theories such extremal solutions
are non-BPS. Geometrically, these `non-BPS-type' solutions
are characterized by null geodesics, or, for multi-centred
solutions, totally geodesic, totally isotropic submanifolds,
where the tangent vectors
do not belong to the eigendistributions of the para-complex
structure. This provides a geometrical
characterization of `non-BPS-type' solutions, which 
applies to supersymmetric and as well non-supersymmetric
theories. A non-trivial charge rotation matrix allows one to
explicitly construct totally geodesic, totally isotropic 
submanifolds starting from the eigendistributions of the 
para-complex structure. We remark that from this point of view
the existence of non-BPS (type) is less generic (or at least less
obvious) than the existence of BPS (type) solutions.

\section{Conclusions}

By dimensional reduction from five to three Euclidean dimensions
we have shown that non-extremal black string solutions 
correspond to space-like geodesi\-cs in the manifold
$S=N \times \mathbb{R}$, where $N\simeq T^*M$ is a para-K\"ahler
manifold which can be identified with 
the cotangent bundle of the manifold $M$ encoding the
couplings of the original five-dimensional theory. Extremal
black string solutions correspond to null geodesics in $N$.
Our construction is not limited to minimal supergravity coupled
to abelian vector multiplets but applies as well to 
Einstein-Maxwell-Scalar theories where all couplings are encoded
by a single homogeneous function.

For BPS-type extremal solutions, where the null geodesics are
contained in the eigendistributions of the para-complex structure,
we can always find explicit solutions where all five-dimensional
scalar fields are independent, with horizon values determined by
the attractor mechanism in terms of the magnetic charges. These
solutions involve $n_V^{(5)}$ real scalars and $n_V^{(5)}+1$ vector
fields and depend on $2n_V^{(5)}+1$ independent integration constants,
namely the values of the scalars at infinity and the magnetic
charges. For supergravity theories we recover the known BPS string
solutions of \cite{Chamseddine:1999qs}. 

Non-extremal solutions and a second type of extremal solutions,
dubbed non-BPS-type solutions can be found {\em explicitly} if
the metric of the scalar manifold admits a non-trivial charge rotation
matrix. The `best case' is provided by diagonal models, where the
metric is diagonal and charge rotation matrices allow one to flip all
charges independently.
For this case we have found explicit non-extremal solutions depending
on $2n_V^{(5)}+2$ independent parameters, which can be taken to
be the values of the scalars at infinity, the magnetic charges and
the non-extremality parameter, and extremal solutions depending
on $2n_V^{(5)} +1$ independent parameters. While the non-BPS-type extremal solutions are of course subject to the attractor mechanism 
we observe a deformed attractor mechanism at work for non-extremal
solutions: while the horizon values of the scalars are no longer determined by the magnetic charges alone, they do not
become independent integration constants. Moreover, the functional
dependence of the horizon values of the scalars was
cast in the form of `horizon charges', both for the inner and
outer horizon.

While diagonal models
constitute a special, non-generic, class of models, this class
contains interesting models, such as the $ST^2$ and $STU$ models
of supergravity and $STU$-like models in non-supersymmetric
theories. These examples were analysed in some detail. 
For non-diagonal models some non-extremal and non-BPS-type extremal
solutions can still be constructed {\em explicitly} if the metric
admits a block decomposition. One important problem left for
future work is to find explicit non-extremal and non-BPS-type 
extremal solutions without the need of a charge rotation matrix
compatible with the metric. Note that the relation 
between non-extremal, non-BPS-type extremal and BPS-type
extremal solutions with particular types of geodesic curve
in $S=N\times \mathbbm{R}$ holds irrespective of whether 
we are able to find solutions explicitly. The distinguished
feature of BPS-type solutions, namely that one can always
find explicit solutions in terms of harmonic functions, corresponds
to the existence of a distinguished class of totally isotropic,
totally geodesic submanifolds associated with the eigendirections
of the para-complex structure. This explains why non-BPS extremal
solutions are harder to find explicitly (unless the metric has special
properties), despite the fact that one might expect 
that one `just needs to flip signs of charges'. While this is true
for `double-extreme' solutions with constant scalar fields, the scalar
equations become in general more complicated because they no
longer decouple.

We finish by pointing out some directions for future research. 
Understanding the precise relation between higher-dimensional 
solutions and geodesic curves and, more generally, totally
geodesic submanifolds of the scalar manifold of a reduced
effective theory should be helpful in analysing the spectrum
of BPS and non-BPS solutions of string theory and M-theory compactifications
in the generic case, where the scalar manifold is not a
symmetric space. One part of the problem is to characterize
submanifolds of the full scalar manifold which are relevant 
for a particular type of higher-dimensional solution, as we 
did here for five-dimensional black strings. Another part is
to investigate which additional conditions one has to impose
on a geodesic curve or totally geodesic submanifold in order
that they lift to regular higher-dimensional solutions. This
determines the number of parameters the higher-dimensional
solution depends on, and has allowed us in this paper 
to recover the attractor mechanism and understand in which sense
it survives in a deformed form for non-extremal solutions. We have
seen that we could also obtain the non-extremal versions of
small black holes by lifting up to four rather than five dimensions.
One might then ask which other types of regular solutions
can be obtained by lifting geodesics with different
boundary conditions. 

For extremal solutions we observed that it is always possible
to give up transverse spherical symmetry and to replace single-centred by multi-centred harmonic functions. Geometrically
such solutions do not correspond to null geodesic curves but 
to totally isotropic totally geodesic submanifolds of $S$.  
Apart from BPS-type solutions, where these submanifolds
are contained within the integrable eigendistributions
of the para-complex structure, we can find explicit non-BPS-type multi-centred solutions whenever a non-trivial charge
rotation matrix exists. Applying such a matrix corresponds
to an overall change of charges at all centres. However, 
in the context of superstring compactifications described
by effective supergravity with symmetric target spaces it
is known that there is a more intricate system of multi-centred
solutions which is not covered by this single operation
\cite{Denef:2000nb,Bates:2003vx,Bossard:2009my,Bossard:2011kz,Ferrara:2012qm}.
%%%here
A deeper understanding of totally geodesic submanifolds
and their relation to multi-centred solutions will be useful
for extending these results to
generic models with non-symmetric target spaces.
%%% more general black strings.

Our work has been restricted to purely magnetic, non-rotating
black strings. More general types of black strings have been
studied in detail for pure five-dimensional supergravity 
in \cite{Compere:2009zh,Compere:2010fm}. Extending these results
to models with vector multiplets would be another possible
extension of the work presented in this paper.

\appendix

%\section{Conventions}
%
%{\bf To be added}
%
%We take the ADM tension $\mathcal{T}$ of a 1-brane in five dimensions to be defined via the asymptotic expansion of $g_{tt}$, as in Appendix B of \cite{Martin:2012bi}:
%\begin{equation}
%g_{tt}\sim 1-\frac{2\mathcal{T}}{\rho} \quad \textrm{as }\rho\rightarrow\infty.
%\end{equation}
%
%{\bf Action on Paul to complete.} 

\section{From Hessian manifolds to para-K\"ahler manifolds
\label{AppHessianPK}}

In this appendix we give a simple self-contained proof that 
the metric on the space $N\simeq T^*M$ appearing in our construction
is a para-K\"ahler metric given that $M$ carries a Hessian metric. 

Let $(M,g)$ be a Hessian manifold. A coordinate-free definition 
can be found in \cite{2008arXiv0811.1658A}. For our purposes we
assume that $M$
is a domain which is covered by a single system of
affine coordinates $w^i$, $i=1,\ldots, n$. We refer to such Hessian manifolds
as Hessian domains. In affine coordinates the metric takes the form
$g=g_{ij}(w) dw^i dw^j$, where $g_{ij}(w) = \partial^2_{i,j} h(w)$
for some function $h(w)$, the Hesse potential\footnote{In the main
part of this paper, the metric coefficients are denoted $\hat{g}_{ij}(w)$.}.
 
%Line element
%$ds_M^2 = g_{ij}(w) dw^i dw^j$, where $g_{ij}(w) = \partial^2_{i,j} h(w)$,
%where $h$ is a Hesse potential. 

Define a new manifold $N=M\times \mathbbm{R}^n$ with coordinates
$(w^i,s_i)$. This manifold can be interpreted as the (trivial)
cotangent bundle of $M$: $N=T^*M$. 
Next, define a pseudo-Riemannian metric on $N$ by
\[
g_N = g_{ij}(w) dw^i dw^j - g^{ij}(w) ds_i ds_j  \;,
\]
where $g^{ij}(w)$ is the inverse matrix of $g_{ij}(w)$. 
The metric $g_N$ obviously has signature $(n,n)$. 
We claim the following statement:
{\em Under the above assumptions, $N=T^*M$ is a para-K\"ahler 
manifold.}\\[2ex]
We refer the reader to \cite{Cortes:2003zd} for the relevant
definitions and theorems on para-complex and para-K\"ahler 
manifolds. We introduce the
frames $F=(\theta_A)=(\partial_{w^i}, \partial_{s_i})$ for $TN$ and 
$F^*=(\theta^A)=(dw^i, ds_i)$ for $T^*N$. The components of $g_N$
are
\[
g_N = g_{AB} \theta^A \theta^B = g_{AB} \theta^A \otimes \theta^B \;,\;\;\;
(g_{AB}) = \left( \begin{array}{cc} 
g & 0 \\
0 & -g^{-1} \\
\end{array} 
\right).
\]
Here we use a block-matrix notation where $g=(g_{ij}(w))$ and
$g^{-1}=(g^{ij}(w))$ are the coefficients of the metric of $M$
and of its inverse with respect to the coordinate system $w^i$. 
Our convention for the symmetrized tensor product is
$\theta^A \theta^B = \frac{1}{2} ( \theta^A \otimes \theta^B + 
\theta^B \otimes \theta^A)$. 

We define an endomorphism
field $J$ on $TN$ 
\[
J= g^{ij} \partial_{w^i} \otimes ds_j + 
g_{ij} \partial_{s_i} \otimes dw^j = J^A_B \theta_A \otimes \theta^B \;.
\]
This acts on the frame $F$ as
\[
J(\partial_{w^i}) = g_{ij} \partial_{s_j}, \;\;\;
J(\partial_{s_i}) = g^{ij} \partial_{w^j} \;.
\]
The components of $J$ with respect to the frame $F$ are
\[
(J^{A}_{B}) = \left( \begin{array}{cc}
0 & g^{-1} \\
g & 0 \\
\end{array} \right) \;.
\]
It follows immediately that 
$J^2 = \mathbbm{1}$. Thus $J$ is an
almost para-complex structure on $N$.
 
The action of $J$ on  $T^*N$ with respect to the dual frame $F^*$
%is given by the transposed matrix:
%\[
%J^* = \left( \begin{array}{cc}
%0 & g \\
%g^_{1} & 0 \\
%\end{array} \right) \;,
%\]
%that 
is
\[
J^* ( dw^i ) = g^{ij} ds_j \;,\;\;\;
J^* ( ds_i) = g_{ij} dw^j \;.
\]
From these expressions it is clear that 
the para-complex structure $J$ acts anti-isometrically on the metric
$g_N = g_{ij} dw^i dw^j - g^{ij} ds_i ds_j$, 
$J^* g = - g$. Therefore $(N,g,J)$ 
is almost para-Hermitian. We define the
fundamental form
\[
\omega = g_N(J\cdot, \cdot)  = \omega_{AB} \theta^A \otimes \theta^B
= \frac{1}{2} \omega_{AB} \theta^A \wedge \theta^B\;.
\]
Our convention for the exterior product is 
$\theta^A \wedge \theta^B = \theta^A \otimes \theta^B - 
\theta^B \otimes \theta^A$.
Evaluate $\omega$ in the frame $F^*$:
\begin{equation}
\label{omega}
\omega =  ds_i \otimes dw^i - dw^i \otimes ds_i =
- dw^i \wedge ds_i, \;\;\;\; 
(\omega_{AB}) = \left( \begin{array}{cc}
0 & -\mathbbm{1} \\
\mathbbm{1} & 0 \\
\end{array} \right) \;.
\end{equation}
%= \underbrace{
%\left( \begin{array}{cc}
%0 & g \\
%g^{-1} & 0 \\
%\end{array} \right)}_{J^T}
%\underbrace{ 
%\left( \begin{array}{cc}
%g & 0 \\
%0 & - g^{-1} \\
%\end{array} \right)}_{g_N} = (g_{CB} J^C_A)\;.
%\]
%As a check, we can read off the components of $\omega$ from 
%(\ref{omega}) or multiply the components of $J$ and $g_N$ 
%in the appropriate way.

We note that $\omega$ 
is the canonical symplectic form on 
$T^*N$. Since $\omega$ is closed, it follows
that $(N,g,J)$ is almost para-K\"ahler. \\[2ex]

It remains to show that $J$ is integrable, which 
is equivalent to showing
that the two eigendistributions are involutive. 
Since $J^2=\mathbbm{1}$, the eigenvalues of $J$ are
$\pm 1$. A basis for the corresponding eigenvectors is
\[
X^i_{\pm} := \frac{1}{\sqrt{2}} \left( 
\partial_{w^i} \pm g_{ij} \partial_{s_j}  \right) \;,
\]
since
\[
J(X^i_\pm) = \frac{1}{\sqrt{2}}
J (\partial_{w^i} \pm g_{ij} \partial_{s_j})
= \frac{1}{\sqrt{2}} g_{ij} \partial_{s_j} \pm g_{ij} g^{jk} \partial_{w^k} 
\]
\[
= \pm \frac{1}{\sqrt{2}} \left( \partial_{w^i} \pm g_{ij} \partial_{s_j}\right)
= \pm X^i_\pm
\;.
\]
The eigenvectors $X^i_\pm$ span the eigendistributions ${\cal D}_\pm$
of $J$. 
We compute the Lie brackets between the eigenvectors:
\[
[X^i_\pm, X^j_\pm] =
\frac{1}{2} [   \partial_{w^i} \pm g_{ij} \partial_{s_j} ,
 \partial_{w^k} \pm g_{kl} \partial_{s_l} ]
= \pm \frac{1}{2} \partial_{w^i} g_{kl} \partial_{s_l} 
\mp \frac{1}{2} \partial_{w^k} g_{ij} \partial_{s_j} = 0\;,
\]
where we used the fact that $\partial_{w^i} g_{jk}$ is totally 
symmetric for the Hessian metric $g_{jk}$.
Thus eigenvectors $X^i_\pm$ belonging to the same
eigendistribution commute, $[X^i_+, X^j_+] = [ X^i_-, X^j_-] = 0$.
This implies that the eigendistributions ${\cal D}_+$ and
${\cal D}_-$  are both involutive, 
therefore $J$ is integrable and $(N,g_N,J)$ is para-K\"ahler. 
This completes the proof. 

For some purposes it is useful to use a frame and co-frame
with respect to which the para-complex structure is diagonal.
Such a frame might be called an `eigenframe' or isotropic
frame (as it is spanned by null vectors). We already saw that 
$F'=(X^i_+, X^i_-)$ is an eigenframe. The associated co-frame is  
\[F'^*=\left(\frac{1}{\sqrt{2}} \left( dw^i + g^{ij} ds_j\right)\;,
\frac{1}{\sqrt{2}} \left( dw^i - g^{ij} ds_j\right) \right).
\]
With respect to these frames the components of the metric
and of the para-complex structure are
\begin{equation}
\label{IsotropicFrame}
(g'_{AB}) = 
\left(\begin{array}{cc}
0 & g \\
g & 0 \\
\end{array} \right) \;,\;\;\;
(J'^A_B) = 
\left(\begin{array}{cc}
1 & 0 \\
0 & -1 \\
\end{array} \right) \;.
\end{equation}
These expressions make manifest that tangent vectors
of the form $(\dot{w}^i,\pm g^{ij}\dot{s_j})$ are isotropic
and moreover are contained in the eigendistributions $\cal{D}_\pm$ of the
para-complex structure. This provides a characterization of
BPS in contrast to non-BPS extremal solutions, which generalizes
to non-supersymmetric theories.

We remark that it is clear that the following more general 
statement is true: {\em The cotangent bundle of a Hessian manifold
carries a natural para-K\"ahler structure.} In other words
one can drop the assumption that the Hessian manifold is a domain
covered by a single affine coordinate system.
This can be shown
by adapting the results of \cite{2008arXiv0811.1658A}, where it was
proven that the tangent bundle of a Hessian manifold carries 
a natural K\"ahler structure. Replacing complex by para-complex
structures amounts to systematically changing certain signs,
see \cite{Cortes:2003zd,Cortes:2009cs}. In addition one has to replace the
tangent bundle by the cotangent bundle using the natural isomorphism
provided by the metric. 

We further remark that a similar situation arises in the case
of the supergravity $r$-map and its generalization to non-supersymmetric
theories. As shown in \cite{Mohaupt:2009iq}, the dimensional 
reduction with respect to time of (not necessarily supersymmetric) 
five-dimensional 
Einstein-Maxwell-Scalar
theories encoded by a homogeneous Hesse potential relates 
Hessian manifolds $(M,g)$ to para-K\"ahler manifolds $(\tilde{N}, 
g_{\tilde{N}})$, where $\tilde{N}$ can be identified with the
tangent bundle $TM$ of $M$. This reduction has been used to
construct the `electric cousins' of the black strings found 
in this paper, see \cite{Mohaupt:2009iq,Mohaupt:2010fk,Mohaupt:2012tu}.

\subsubsection*{Acknowledgements}

We thank Vicente Cort\'{e}s and Owen Vaughan for useful
discussions.
The work of T.M. is supported in part by STFC  
grants ST/G00062X/1 and ST/J000493/1.
The work of P.D. is supported by STFC studentship ST/1505805/1. 
%We thank Vicente Cort\'es, Bernard de Wit and Dieter Van den Bleeken
%for useful discussions. 

%\bibliography{LitbankW}	
%\bibliography{BS}
%\bibliography{mybib}{}

\begin{thebibliography}{10}

\bibitem{Ferrara:1995ih}
S. Ferrara, R. Kallosh and A. Strominger,
\newblock Phys. Rev. D52 (1995) 5412, hep-th/9508072.

\bibitem{Strominger:1996sh}
A. Strominger and C. Vafa,
\newblock Phys. Lett. B379 (1996) 99, hep-th/9601029.

\bibitem{Maldacena:1997de}
J.M. Maldacena, A. Strominger and E. Witten,
\newblock JHEP 9712 (1997) 002, hep-th/9711053.

\bibitem{Vafa:1997gr}
C. Vafa,
\newblock Adv.Theor.Math.Phys. 2 (1998) 207, hep-th/9711067.

\bibitem{LopesCardoso:1998wt}
G. Lopes~Cardoso, B. de~Wit and T. Mohaupt,
\newblock Phys.Lett. B451 (1999) 309, hep-th/9812082.

\bibitem{Ferrara:1997tw}
S. Ferrara, G.W. Gibbons and R. Kallosh,
\newblock Nucl. Phys. B500 (1997) 75, hep-th/9702103.

\bibitem{Goldstein:2005hq}
K. Goldstein et~al.,
\newblock Phys. Rev. D72 (2005) 124021, hep-th/0507096.

\bibitem{Tripathy:2005qp}
P.K. Tripathy and S.P. Trivedi,
\newblock JHEP 03 (2006) 022, hep-th/0511117.

\bibitem{Sen:2005wa}
A. Sen,
\newblock JHEP 0509 (2005) 038, hep-th/0506177.

\bibitem{MyersPerry}
R. Myers and M. Perry,
\newblock Ann. Phys. 172 (1986) 304.

\bibitem{Gibbons:1987ps}
G.W. Gibbons and K.I. Maeda,
\newblock Nucl. Phys. B298 (1988) 741.

\bibitem{Garfinkle:1990qj}
D. Garfinkle, G.T. Horowitz and A. Strominger,
\newblock Phys. Rev. D43 (1991) 3140.

\bibitem{Horowitz:1991cd}
G.T. Horowitz and A. Strominger,
\newblock Nucl. Phys. B360 (1991) 197.

\bibitem{Cvetic:1995dn}
M. Cvetic and D. Youm,
\newblock (1995), hep-th/9508058.

\bibitem{Cvetic:1995kv}
M. Cvetic and D. Youm,
\newblock Nucl. Phys. B472 (1996) 249, hep-th/9512127.

\bibitem{Lu:2003iv}
H. Lu, C.N. Pope and J.F. Vazquez-Poritz,
\newblock Nucl. Phys. B709 (2005) 47, hep-th/0307001.

\bibitem{Miller:2006ay}
C.M. Miller, K. Schalm and E.J. Weinberg,
\newblock Phys. Rev. D76 (2007) 044001, hep-th/0612308.

\bibitem{Garousi:2007zb}
M.R. Garousi and A. Ghodsi,
\newblock JHEP 05 (2007) 043, hep-th/0703260.

\bibitem{Andrianopoli:2007gt}
L. Andrianopoli et~al.,
\newblock JHEP 11 (2007) 032, 0706.0712.

\bibitem{Janssen:2007rc}
B. Janssen et~al.,
\newblock JHEP 04 (2008) 007, 0712.2808.

\bibitem{Cardoso:2008gm}
G.L. Cardoso and V. Grass,
\newblock Nucl. Phys. B803 (2008) 209, 0803.2819.

\bibitem{Perz:2008kh}
J. Perz et~al.,
\newblock JHEP 03 (2009) 150, 0810.1528.

\bibitem{Meessen:2011bd}
P. Meessen and T. Ortin,
\newblock (2011), 1107.5454.

\bibitem{Martin:2012bi}
A.d.A. Martin, T. Ortin and C.S. Shahbazi,
\newblock (2012), 1203.0260.

\bibitem{Meessen:2012su}
P. Meessen et~al.,
\newblock (2012), 1204.0507.

\bibitem{Cortes:2003zd}
V. Cort{\'e}s et~al.,
\newblock JHEP 03 (2004) 028, hep-th/0312001.

\bibitem{Cortes:2005uq}
V. Cort{\'e}s et~al.,
\newblock JHEP 06 (2005) 025, hep-th/0503094.

\bibitem{Cortes:2009cs}
V. Cort{\'e}s and T. Mohaupt,
\newblock JHEP 07 (2009) 066, 0905.2844.

\bibitem{Euc4}
V. Cort{\'e}s et~al.,
\newblock {Special Geometry of Euclidean Supersymmetry IV: hypermultiplets and
  local c-maps},
\newblock e-print to appear.

\bibitem{Mohaupt:2009iq}
T. Mohaupt and K. Waite,
\newblock JHEP 10 (2009) 058, 0906.3451.

\bibitem{Mohaupt:2010fk}
T. Mohaupt and O. Vaughan,
\newblock Class. Quant. Grav. 27 (2010) 235008, 1006.3439.

\bibitem{Mohaupt:2012tu}
T. Mohaupt and O. Vaughan,
\newblock (2012), 1208.4302.

\bibitem{Mohaupt:2011aa}
T. Mohaupt and O. Vaughan,
\newblock JHEP 1207 (2012) 163, 1112.2876.

\bibitem{Gunaydin:1983bi}
M. Gunaydin, G. Sierra and P.K. Townsend,
\newblock Nucl. Phys. B242 (1984) 244.

\bibitem{Chamseddine:1999qs}
A.H. Chamseddine and W.A. Sabra,
\newblock Phys. Lett. B460 (1999) 63, hep-th/9903046.

\bibitem{Breitenlohner:1987dg}
P. Breitenlohner, D. Maison and G.W. Gibbons,
\newblock Commun. Math. Phys. 120 (1988) 295.

\bibitem{G2}
V. Cort\'es, P. Dempster and T. Mohaupt,
\newblock Timelike reductions of five-dimensional supergravity,
\newblock to appear.

\bibitem{Mayer:2003zk}
C. Mayer and T. Mohaupt,
\newblock Class. Quant. Grav. 21 (2004) 1879, hep-th/0312008.

\bibitem{Antoniadis:1995vz}
I. Antoniadis, S. Ferrara and T.R. Taylor,
\newblock Nucl. Phys. B460 (1996) 489, hep-th/9511108.

\bibitem{Chou:1997ba}
A.S. Chou et~al.,
\newblock Nucl. Phys. B508 (1997) 147, hep-th/9704142.

\bibitem{Kallosh:2000rn}
R. Kallosh, T. Mohaupt and M. Shmakova,
\newblock J.Math.Phys. 42 (2001) 3071, hep-th/0010271.

\bibitem{Ceresole:2007wx}
A. Ceresole and G. Dall'Agata,
\newblock JHEP 03 (2007) 110, hep-th/0702088.

\bibitem{LopesCardoso:2007ky}
G. Lopes~Cardoso et~al.,
\newblock JHEP 10 (2007) 063, 0706.3373.

\bibitem{LopesCardoso:2000qm}
G. Lopes~Cardoso et~al.,
\newblock JHEP 12 (2000) 019, hep-th/0009234.

\bibitem{Dabholkar:2004dq}
A. Dabholkar, R. Kallosh and A. Maloney,
\newblock JHEP 0412 (2004) 059, hep-th/0410076.

\bibitem{2008arXiv0811.1658A}
D.V. {Alekseevsky} and V. {Cort{\'e}s},
\newblock Commun. Math. Phys. 291 (2009) 579, 0811.1658.

\bibitem{Denef:2000nb}
F. Denef,
\newblock JHEP 08 (2000) 050, hep-th/0005049.

\bibitem{Bates:2003vx}
B. Bates and F. Denef,
\newblock JHEP 1111 (2011) 127, hep-th/0304094.

\bibitem{Bossard:2009my}
G. Bossard and H. Nicolai,
\newblock Gen.Rel.Grav. 42 (2010) 509, 0906.1987.

\bibitem{Bossard:2011kz}
G. Bossard and C. Ruef,
\newblock Gen.Rel.Grav. 44 (2012) 21, 1106.5806.

\bibitem{Ferrara:2012qm}
S. Ferrara et~al.,
\newblock (2012), 1211.3262.

\bibitem{Compere:2009zh}
G. Comp\`ere et~al.,
\newblock Class.Quant.Grav. 26 (2009) 125016, 0903.1645.

\bibitem{Compere:2010fm}
G. Comp\`ere et~al.,
\newblock JHEP 1011 (2010) 133, 1006.5464.

\end{thebibliography}
%\bibliographystyle{plain}
%\bibliographystyle{hplain}
%\bibliographystyle{hep}
%\bibliographystyle{h-elsevier}
%\bibliographystyle{kp}

\end{document}